\pdfoutput=1
\documentclass[twocolumn]{aastex631}

\usepackage{amsmath}
\usepackage{booktabs}
\usepackage{graphicx}
\usepackage{multirow}
\graphicspath{{figures/}}

\newcommand{\BlueRedshift}{2.1745}
\newcommand{\RedRedshift}{2.1965}
\newcommand{\PeakSeparation}{2073}
\newcommand{\PeakSeparationLo}{1992}
\newcommand{\PeakSeparationHi}{2172}

\newcommand{\ProjectedSeparation}{0.49}

\newcommand{\CentroidPValue}{0.641}

\newcommand{\RefRatioOfficial}{2.95}

\newcommand{\RefRatioEmpirical}{1.04}

\newcommand{\RefBlueGapper}{170}
\newcommand{\RefBlueGapperLo}{114}
\newcommand{\RefBlueGapperHi}{200}

\newcommand{\RefRedGapper}{225}
\newcommand{\RefRedGapperLo}{151}
\newcommand{\RefRedGapperHi}{257}

\newcommand{\RefNInsideProjectedMid}{19}

\newcommand{\RefPairFisherP}{0.035}

\newcommand{\RefMinBindMass}{6.4}

\newcommand{\RefProjSepLo}{0.33}
\newcommand{\RefProjSepHi}{1.31}

\newcommand{\RefMassBlueGapper}{1.8}

\newcommand{\RefMassRedGapper}{4.2}

\newcommand{\RefMassAllGapper}{7.5}
\newcommand{\RefMassAllGapperLo}{4.3}
\newcommand{\RefMassAllGapperHi}{9.4}

\newcommand{\PaperZRef}{2.1855}

\newcommand{\PaperRespN}{27}
\newcommand{\PaperRespDchi}{22.3}

\newcommand{\PaperClassABlue}{5}
\newcommand{\PaperClassARed}{8}

\newcommand{\PhaseRtaHigh}{4.3}

\newcommand{\QsoNBeyondTurnaroundHigh}{4}

\newcommand{\BroadNinetyNTwo}{-0.11}
\newcommand{\BroadNinetyNTwoErr}{0.04}

\newcommand{\MergeBoundDCMpc}{37}
\newcommand{\MergeBoundDelta}{0.65}

\newcommand{\MergeBoundSigmaLin}{0.13}
\newcommand{\MergeBoundNSigma}{3.4}
\newcommand{\MergeBoundPRandom}{$4\times10^{-4}$}
\newcommand{\MergeBoundDeltaGal}{1.6}
\newcommand{\MergeBoundRadiusArcmin}{23}
\newcommand{\MergeBoundDzHalf}{0.027}
\newcommand{\MergeMergedDCMpc}{45.5}
\newcommand{\MergeMergedDelta}{1.23}

\newcommand{\MergeMergedSigmaLin}{0.10}
\newcommand{\MergeMergedNSigma}{6.0}
\newcommand{\MergeMergedPRandom}{$9\times10^{-10}$}
\newcommand{\MergeMergedDeltaGal}{3.0}
\newcommand{\MergeMergedRadiusArcmin}{28}
\newcommand{\MergeMergedDzHalf}{0.034}

\newcommand{\MergeMassAtBound}{1.5}

\newcommand{\QsoNMosfire}{3}
\newcommand{\QsoNCo}{3}
\newcommand{\QsoCoSdssOffsetLo}{$-$422}
\newcommand{\QsoCoSdssOffsetHi}{+745}
\newcommand{\QsoMosfireMinusCoQOne}{$-$43}
\newcommand{\QsoNSdss}{5}

\newcommand{\QsoBlueSideMeanV}{$-$1466}
\newcommand{\QsoRedSideMeanV}{+729}

\newcommand{\QsoGapLoV}{$-$843}
\newcommand{\QsoGapHiV}{+599}
\newcommand{\QsoGapP}{0.049}

\newcommand{\QsoSep}{5.2}
\newcommand{\QsoSepP}{0.009}

\newcommand{\QsoDecOffset}{5.1}

\newcommand{\QsoSepPFracSig}{95}

\newcommand{\QsoSdssErr}{500}
\newcommand{\QsoSdssOffsetLo}{$-$379}
\newcommand{\QsoSdssOffsetHi}{+394}
\newcommand{\QsoSdssOffsetQTwo}{+1333}
\newcommand{\QsoBroadPeakMax}{433}
\newcommand{\QsoRMaxSdss}{8.4}

\newcommand{\QsoBlueSideMinusComponent}{429}
\newcommand{\QsoRedSideMinusComponent}{307}

\newcommand{\MsepDeltaTurn}{4.55}
\newcommand{\MsepDZero}{30}

\newcommand{\MsepDHalf}{35}
\newcommand{\MsepDOne}{42}
\newcommand{\MsepDTwo}{62}

\newcommand{\MsepApproachMin}{1.0}
\newcommand{\MsepApproachMinD}{0.7}

\newcommand{\GapObs}{1268}
\newcommand{\GapLo}{$-$691}
\newcommand{\GapHi}{+577}
\newcommand{\GapSD}{1300}
\newcommand{\GapExpectedNone}{9}

\newcommand{\GapNNone}{78}
\newcommand{\GapNEmp}{23}
\newcommand{\GapNOff}{3}
\newcommand{\GapNSim}{200{,}000}

\newcommand{\SuccPhotDet}{3{,}222}

\newcommand{\SuccPhotN}{117}

\newcommand{\SuccPhotHAE}{105}

\newcommand{\SuccSigmaMin}{3}
\newcommand{\SuccEWRestMin}{25}

\newcommand{\SuccTargetsN}{54}

\newcommand{\SuccMembersOutN}{eight}

\newcommand{\SuccCandN}{17}

\newcommand{\SuccCandHa}{16}

\newcommand{\SuccCandSuccess}{94}
\newcommand{\SuccCandSuccessLo}{86}
\newcommand{\SuccCandSuccessHi}{98}

\newcommand{\SuccFillerN}{37}

\newcommand{\ChAreaJoint}{105}
\newcommand{\ChDepthNB}{23.21}
\newcommand{\ChDepthKs}{23.50}

\newcommand{\ChPsfNBArcsec}{0\farcs61}
\newcommand{\ChPsfKsArcsec}{0\farcs57}

\newcommand{\PosEmpFWHM}{351}

\newcommand{\PosAvgFWHM}{282}

\newcommand{\PosZRedEdgeMin}{2.1950}
\newcommand{\PosZRedEdgeMax}{2.2002}
\newcommand{\PosTRedMin}{34}
\newcommand{\PosTRedMax}{96}
\newcommand{\PosTRedMedian}{68}

\newcommand{\PosDeltaBIC}{15.4}
\newcommand{\PosDeltaBICMin}{14.5}

\newcommand{\PosRatio}{2.41}
\newcommand{\PosRatioLo}{1.56}
\newcommand{\PosRatioHi}{3.85}

\newcommand{\PosDispBlue}{171}

\newcommand{\PosDispRed}{252}

\newcommand{\GapNPos}{3}

\newcommand{\GapNSimCases}{200{,}000}
\newcommand{\GapOHPercent}{7}

\newcommand{\GapNShapeMaxPos}{42}

\newcommand{\GroupLhaP}{0.69}

\newcommand{\GroupNTwoP}{0.25}

\newcommand{\ExtentAgeZRef}{3.0}

\newcommand{\ExtentQsoCoRange}{167}

\newcommand{\ExtentRrmsBlue}{2.0}
\newcommand{\ExtentRrmsBlueLo}{1.6}
\newcommand{\ExtentRrmsBlueHi}{2.1}
\newcommand{\ExtentPairMaxBlue}{5.6}
\newcommand{\ExtentTcrossBlue}{11}
\newcommand{\ExtentTcrossBlueLo}{9}
\newcommand{\ExtentTcrossBlueHi}{16}
\newcommand{\ExtentRvirBlue}{0.12}

\newcommand{\ExtentLosDepthBlue}{0.8}
\newcommand{\ExtentRatioBlue}{17}
\newcommand{\ExtentTcrossThreeDBlue}{6.6}

\newcommand{\ExtentReachBlue}{0.52}

\newcommand{\ExtentRrmsRed}{1.7}
\newcommand{\ExtentRrmsRedLo}{1.4}
\newcommand{\ExtentRrmsRedHi}{1.9}
\newcommand{\ExtentPairMaxRed}{4.8}
\newcommand{\ExtentTcrossRed}{8}
\newcommand{\ExtentTcrossRedLo}{6}
\newcommand{\ExtentTcrossRedHi}{11}
\newcommand{\ExtentRvirRed}{0.15}

\newcommand{\ExtentLosDepthRed}{1.0}
\newcommand{\ExtentRatioRed}{11}
\newcommand{\ExtentTcrossThreeDRed}{4.3}

\newcommand{\ExtentReachRed}{0.69}
\newcommand{\ExtentSigmaBoth}{1052}

\newcommand{\ExtentRrmsBoth}{1.9}

\newcommand{\AssocQsoN}{three}
\newcommand{\AssocQsoVMin}{+1749}
\newcommand{\AssocQsoVMax}{+3514}
\newcommand{\AssocQsoBeyondRedMin}{712}
\newcommand{\AssocQsoBeyondRedMax}{2478}

\newcommand{\AssocQsoLinkLength}{21}

\newcommand{\QsoDesiCoOffsetLo}{$-$432}
\newcommand{\QsoDesiCoOffsetHi}{+1695}

\newcommand{\SensFractionMin}{1}
\newcommand{\SensFractionMax}{10}
\newcommand{\SensDvSel}{2171}

\newcommand{\SensBoundDeltaSel}{0.64}

\newcommand{\SensMergedDeltaSel}{1.23}

\newcommand{\SensQsoGapQEightDesi}{1369}
\newcommand{\SensQsoGapPQEightDesi}{0.070}
\newcommand{\SensQsoSepPQEightDesi}{0.009}
\newcommand{\SensQsoGapPLooMax}{0.073}
\newcommand{\SensQsoGapLooMin}{1441}
\newcommand{\SensQsoSepPLooMin}{0.006}
\newcommand{\SensQsoSepPLooMax}{0.031}

\newcommand{\ChefThreeDCore}{40}
\newcommand{\ChefThreeDWide}{34}

\newcommand{\ChefThreeDmCore}{0.9}
\newcommand{\ChefThreeDmCoreLo}{0.6}
\newcommand{\ChefThreeDmCoreHi}{1.3}

\newcommand{\ChefThreeDmWide}{0.4}
\newcommand{\ChefThreeDmWideLo}{0.2}
\newcommand{\ChefThreeDmWideHi}{0.8}

\newcommand{\InfallHalfMass}{1.3}
\newcommand{\InfallHalfD}{0.4}

\newcommand{\Ha}{H$\alpha$}
\newcommand{\Hb}{H$\beta$}
\newcommand{\NII}{[N~{\sc ii}]}
\newcommand{\SII}{[S~{\sc ii}]}
\newcommand{\OIII}{[O~{\sc iii}]}

\newcommand{\numword}[1]{\ifcase#1 zero\or one\or two\or three\or four\or five\or six\or seven\or eight\or nine\else #1\fi}
\newcommand{\Numword}[1]{\ifcase#1 Zero\or One\or Two\or Three\or Four\or Five\or Six\or Seven\or Eight\or Nine\else #1\fi}
\newcommand{\kms}{\mathrm{km~s^{-1}}}

\newcommand{\Msun}{M_\odot}

\shorttitle{CHEF II. Two cold velocity groups in a quasar overdensity}
\shortauthors{Liang et al.}

\begin{document}

\title{The \textit{Cosmic Himalayas} Extended Field (CHEF). II.\ H$\alpha$ Kinematics of the Densest Quasar Overdensity at $z=2.2$: Two Cold Velocity Groups Separated by 2000~km~s$^{-1}$}

\correspondingauthor{Yongming Liang}
\email{ymliang@icrr.u-tokyo.ac.jp}
\author[0000-0002-2725-302X]{Yongming Liang}
\affiliation{Institute for Cosmic Ray Research, The University of Tokyo, 5-1-5 Kashiwanoha, Kashiwa, Chiba 277-8582, Japan}
\affiliation{National Astronomical Observatory of Japan, 2-21-1 Osawa, Mitaka, Tokyo 181-8588, Japan}
\author[0000-0002-1049-6658]{Masami Ouchi}
\affiliation{National Astronomical Observatory of Japan, 2-21-1 Osawa, Mitaka, Tokyo 181-8588, Japan}
\affiliation{Institute for Cosmic Ray Research, The University of Tokyo, 5-1-5 Kashiwanoha, Kashiwa, Chiba 277-8582, Japan}
\author[0000-0001-5804-1428]{Sebastiano Cantalupo}
\affiliation{Department of Physics, University of Milan Bicocca, Piazza della Scienza 3, I-20126 Milano, Italy}
\affiliation{INAF--Osservatorio Astronomico di Trieste, Via G. B. Tiepolo 11, 34143 Trieste, Italy}
\author[0000-0001-9687-4973]{Naveen Reddy}
\affiliation{Department of Physics and Astronomy, University of California, Riverside, USA}
\author[0000-0002-4834-7260]{Charles Steidel}
\affiliation{Cahill Center for Astronomy and Astrophysics, California Institute of Technology, USA}
\author[0000-0001-7457-8487]{Kentaro Nagamine}
\affiliation{Department of Earth and Space Science, Osaka University, 1-1 Machikaneyama, Toyonaka, Osaka 560-0043, Japan}
\author[0009-0005-1273-0742]{Yuto Kuwayama}
\affiliation{Department of Earth and Space Science, Osaka University, 1-1 Machikaneyama, Toyonaka, Osaka 560-0043, Japan}
\author[0000-0002-1319-3433]{Hidenobu Yajima}
\affiliation{Center for Computational Sciences, University of Tsukuba, 1-1-1 Tennodai, Tsukuba, Ibaraki 305-8577, Japan}
\author[0000-0003-4442-2750]{Rhythm Shimakawa}
\affiliation{Waseda Institute for Advanced Study (WIAS), Waseda University, 1-21-1 Nishi-Waseda, Shinjuku, Tokyo 169-0051, Japan}
\author[0000-0003-3954-4219]{Nobunari Kashikawa}
\affiliation{Department of Astronomy, School of Science, The University of Tokyo, 7-3-1 Hongo, Bunkyo-ku, Tokyo, 113-0033, Japan}
\author[0000-0002-2993-1576]{Tadayuki Kodama}
\affiliation{Astronomical Institute, Tohoku University, 6-3 Aramaki, Aoba-ku, Sendai, Miyagi 980-8578, Japan}
\author[0000-0002-5011-5178]{Masayuki Tanaka}
\affiliation{National Astronomical Observatory of Japan, 2-21-1 Osawa, Mitaka, Tokyo 181-8588, Japan}
\author[0000-0002-4314-5686]{Dongdong Shi}
\affiliation{Center for Fundamental Physics, School of Mechanics and Optoelectronic Physics, Anhui University of Science and Technology, Huainan, Anhui 232001, China}
\author[0000-0002-9453-0381]{Kei Ito}
\affiliation{Cosmic Dawn Center (DAWN), Denmark}
\affiliation{DTU Space, Technical University of Denmark, Elektrovej 327, DK2800 Kgs. Lyngby, Denmark}
\author[0000-0003-3310-0131]{Xiaohui Fan}
\affiliation{Steward Observatory, University of Arizona, 933 N. Cherry Avenue, Tucson, AZ 85721, USA}
\author[0000-0001-8467-6478]{Zheng Cai}
\affiliation{Department of Astronomy, Tsinghua University, Beijing 100084, China}
\author[0000-0001-6251-649X]{Mingyu Li}
\affiliation{Department of Astronomy, Tsinghua University, Beijing 100084, China}
\author[0000-0002-9373-3865]{Xin Wang}
\affiliation{School of Astronomy and Space Science, University of Chinese Academy of Sciences (UCAS), Beijing 100049, China}
\affiliation{National Astronomical Observatories, Chinese Academy of Sciences, Beijing 100101, China}
\affiliation{Institute for Frontiers in Astronomy and Astrophysics, Beijing Normal University, Beijing 102206, China}
\author[0000-0002-9593-8274]{Weichen Wang}
\affiliation{Chinese Academy of Sciences South America Center for Astronomy (CASSACA), NAOC, Beijing 100101, China}
\affiliation{Centro Conjunto China-Chile de Astronomía (CCJCA), Departamento de Astronomía, Universidad de Chile, Santiago, Chile}
\author[0000-0002-8863-888X]{Andrea Travascio}
\affiliation{INAF--Osservatorio Astronomico di Trieste, Via G. B. Tiepolo 11, 34143 Trieste, Italy}
\affiliation{Department of Physics, University of Milan Bicocca, Piazza della Scienza 3, I-20126 Milano, Italy}
\author[0009-0004-4332-9225]{Tomokazu Kiyota}
\affiliation{Astronomical Science Program, Graduate Institute for Advanced Studies, SOKENDAI, 2-21-1 Osawa, Mitaka, Tokyo 181-8588, Japan}
\affiliation{National Astronomical Observatory of Japan, 2-21-1 Osawa, Mitaka, Tokyo 181-8588, Japan}

\begin{abstract}
The 11 quasars at $z=2.16$--2.20 of the Cosmic Himalayas, the densest quasar overdensity in the Sloan Digital Sky Survey, could mark one massive halo, several halos along the line of sight, or part of a filament.
Keck/MOSFIRE spectroscopy of narrowband-selected galaxies in its central \ChAreaJoint~arcmin$^2$ confirms 24 H$\alpha$ emitters.
Their redshifts split into two velocity groups at $z=\BlueRedshift$ and $\RedRedshift$, \PeakSeparation~km~s$^{-1}$ apart, with not one emitter in the \GapObs~km~s$^{-1}$ between them.
The 11 quasars, which reach \QsoRMaxSdss~proper Mpc from the core, split at the same velocities, with the red side \QsoDecOffset~proper Mpc north of the blue side.
Each group is kinematically cold, with a dispersion of \RefBlueGapper\ or \RefRedGapper~km~s$^{-1}$, yet its members lie up to \ExtentPairMaxBlue~proper Mpc apart on the sky, so that its crossing time exceeds the age of the Universe.
Each group, only about 1~proper Mpc deep in Hubble flow, could be a sheet or filament seen side-on or an uncollapsed region of a protocluster.
In a spherical model, the mean matter overdensity $\delta$ around both groups sets their separation, \MsepDZero--\MsepDOne\ comoving Mpc for $\delta=0$--1, and their fate, from never meeting ($\delta<\MergeBoundDelta$) to merging before $z=0$ ($\delta>\MergeMergedDelta$), while the core alone does not exclude one structure larger than the masks.
A one-square-degree survey with a wide-field spectrograph such as the Subaru Prime Focus Spectrograph can map the groups beyond the masks and measure $\delta$.
\end{abstract}


\section{Introduction}\label{sec:introduction}

Luminous quasars at $z\approx2$--3 cluster like dark matter halos of $\sim10^{12}~h^{-1}~\Msun$ \citep{White2012,Eftekharzadeh2015}.
Yet their clustering below $\sim40~h^{-1}$~kpc exceeds the extrapolation from larger scales \citep{Hennawi2006}, four quasars lie within one giant Ly$\alpha$ nebula in a galaxy overdensity at $z\approx2$ \citep{Hennawi2015}, and protoclusters at these redshifts host more active nuclei than the field \citep{Lehmer2009,Tozzi2022}.
Merger-driven models place quasars in small galaxy groups, where mergers of gas-rich galaxies are most efficient \citep{Cavaliere2000,Hopkins2008}, and in simulations cold gas that flows along the filaments of the cosmic web feeds galaxies and their black holes \citep{Dekel2009,DiMatteo2012,Dubois2012}.
Where many quasars appear together, the structure that hosts them shows in which environment, a massive halo, a small galaxy group, or a filament, several quasars are active at once.

The Cosmic Himalayas \citep[CH;][]{Liang2025} is the densest quasar overdensity in the $\sim10{,}000$~deg$^2$ of the Sloan Digital Sky Survey (SDSS), with 11 quasars at $z=2.16$--$2.20$ and a $17\sigma$ excess.
Such a concentration is rare but occurs in the CROCODILE $\Lambda$CDM simulation \citep{Kuwayama2026}.
If the quasars marked the most massive halo of the field, they would coincide with the density peaks of galaxies and gas.
Instead, the two density peaks of Ly$\alpha$ emitters lie about $25~h^{-1}$ comoving Mpc (cMpc) to the northwest and southeast of the quasar density peak, which also coincides with no peak of intergalactic H~{\sc i} absorption \citep{Liang2025}.
Ly$\alpha$ emitters are rarer than continuum-selected galaxies close to luminous quasars, however \citep{Kashikawa2007,Suzuki2025}, so this offset alone does not show where the mass lies.
Luminous quasars at $z\approx3.8$ reside in environments like those of star-forming galaxies, and the most luminous of them tend to avoid the most overdense regions \citep{Uchiyama2018}.
It is unclear whether the CH quasars mark one massive halo, several halos along the line of sight, or part of a filament.
Several protoclusters at $z>2$ contain two velocity groups about 1500--2000~$\kms$ apart, as SSA22 \citep{Topping2016}, the Spiderweb \citep{Kuiper2011}, and BOSS1244 \citep{Shi2021} do, interpreted either as halos falling into each other \citep{Kuiper2011} or as separate structures along the line of sight \citep{Topping2018}.
In the Spiderweb, a larger sample of \Ha\ emitters shows filaments that cross the center instead \citep{Shimakawa2014}.
\Ha\ traces the systemic redshifts of star-forming galaxies at $z\approx2$ far more precisely than such velocity differences, so the velocities of \Ha\ emitters around the CH quasars can distinguish between these cases.

We use the quasar overdensity as a tracer of the structures around it and present Keck/MOSFIRE spectroscopy of \Ha\ emitters (HAEs), selected with the NB2083 narrowband filter at the redshift of the quasars, in the CH core.
The core, the area of our Subaru/MOIRCS imaging around the quasar density peak, covers \ChAreaJoint~arcmin$^2$ and spans about 8 proper Mpc along its longest axis ($1'$ is 0.51 proper Mpc at $z=2.19$).
The HAE redshifts form two kinematically cold but extended velocity groups (hereafter groups), at whose velocities the quasars divide (Section~\ref{sec:results}), and Section~\ref{sec:nature} relates their configuration and fate to the mean overdensity of the region around them.
Section~\ref{sec:quasars} describes how a wide-field spectroscopic survey of about one square degree can tell these configurations apart.
We adopt the \citet{Planck2018} cosmology, and distances are proper unless stated as comoving.
Velocities are rest-frame offsets from $z_{\rm ref}=\PaperZRef$, the mean of the two group centers, with $\Delta v=c~(z-z_{\rm ref})/(1+z_{\rm ref})$.

\section{Target selection and spectroscopy}\label{sec:data}\label{sec:selection}

The targets come from CHEF~I (Y.~Liang et al., in preparation), the master catalog of the field, built on our MOIRCS imaging (Appendix~\ref{app:imaging}).
The NB2083 filter (central wavelength 2.084~\micron, FWHM 281~\AA\ in the manufacturer's curve for 5\degr\ incidence, hereafter the official curve) selects \Ha\ at $z=2.152$--$2.195$.
Its passband moves blueward by up to 110~\AA\ toward the corners of the MOIRCS field, so CHEF~I assigns each source the passband at its position.

\begin{figure*}[t]
\centering
\includegraphics[width=\textwidth]{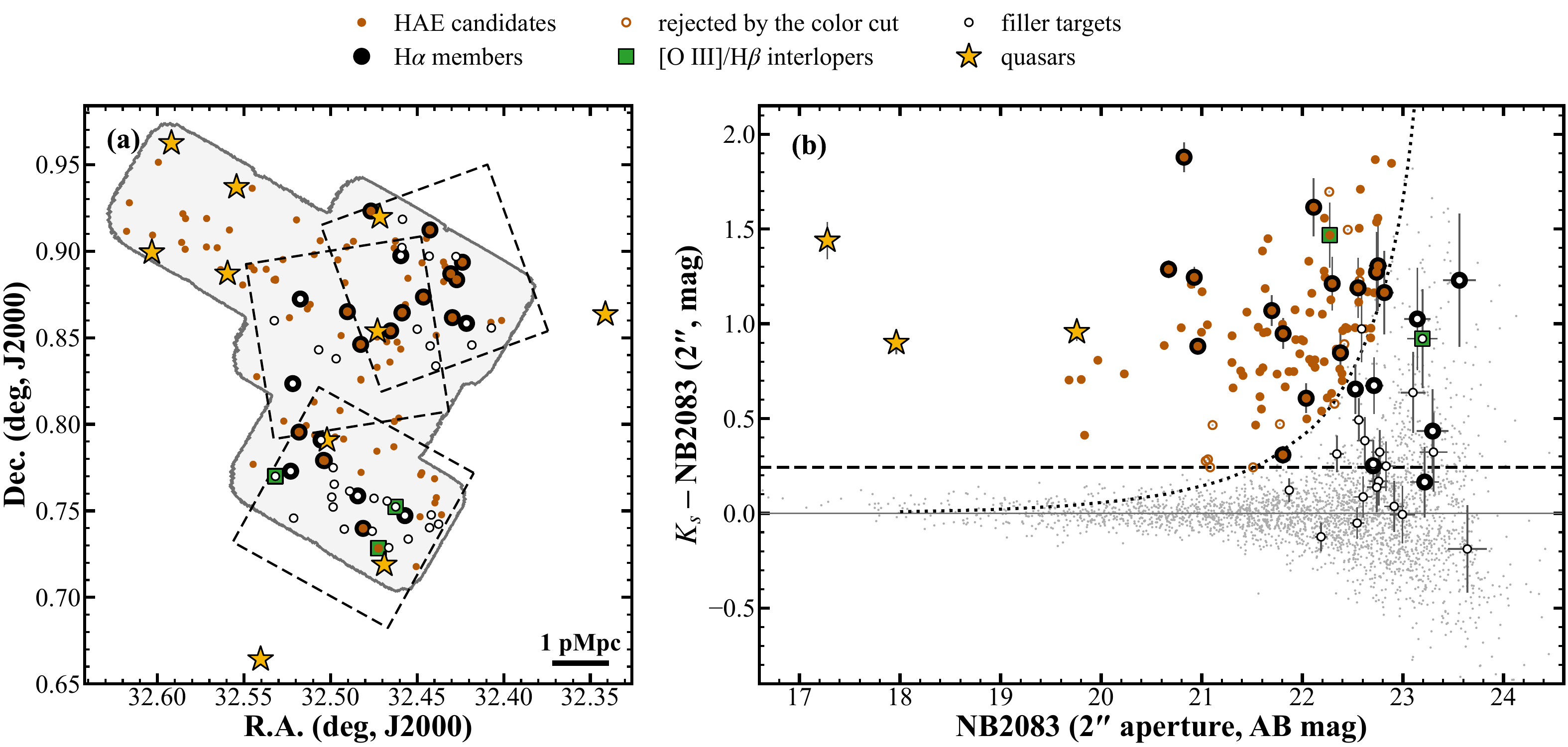}
\caption{The CH core and the selection of the targets.
Each target is a dot for its selection (orange: HAE candidate, white: filler target) on a symbol for its outcome (black circle: \Ha\ member, green square: \OIII\ or \Hb\ interloper, none: no confirmed line).
Gold stars: the CH quasars, of which the masks targeted CH-Q01, CH-Q02, CH-Q05, and CH-Q07 (CH-Q06 lies 8.4~pMpc to the north, outside panel a).
(a) The sky, east to the left, with the core (gray, the MOIRCS imaging), the MOSFIRE masks (dashed), the \SuccPhotHAE\ HAE candidates (orange dots), and a bar of 1 proper Mpc (pMpc).
(b) $K_s-{\rm NB2083}$ color against NB2083 magnitude in $2''$ apertures, for valid NB2083 detections (gray) and targets ($1\sigma$ errors).
Narrowband-excess sources lie above the dashed line (equivalent width of \SuccEWRestMin~\AA) and the dotted curve ($\Sigma=3$ at the median noise).
Thin line: zero color.
Open orange circles: rejected by the color cut.
Not shown in (b): targets without a valid NB2083 detection, such as the saturated CH-Q01, and six sources with colors above 2.15.}
\label{fig:field}
\end{figure*}

HAE candidates are the narrowband-excess sources of CHEF~I (excess significance $\Sigma\geq\SuccSigmaMin$ and a rest-frame \Ha+\NII\ equivalent width of at least \SuccEWRestMin~\AA, Figure~\ref{fig:field}b) that pass its color cut against lower-redshift interlopers.
Of the \SuccPhotDet\ NB2083 detections, \SuccPhotN\ are narrowband-excess sources and \SuccPhotHAE\ are HAE candidates.

Three Keck~I/MOSFIRE masks \citep{McLean2012} with $1''$ slits observed the $K$ band for 5.4~hr on source in total, at $R\approx2700$ (Appendix~\ref{app:imaging}).
They targeted four of the eight CH quasars in the core and \SuccTargetsN\ galaxies of a preliminary MOIRCS catalog.
In the final catalog \SuccCandN\ of these galaxies are HAE candidates, and the remaining \SuccFillerN\ non-quasar targets are filler targets (Figure~\ref{fig:field}).
The spectra confirm 24 \Ha\ emitters at $z=2.15$--2.20 (hereafter members) and three \OIII\ or \Hb\ emitters at $z\approx3.2$, and show three tentative lines that we do not use (Appendix~\ref{app:atlas}).
Class A members show \Ha\ with \NII~$\lambda6585$ or \SII\ at the same redshift.
Class B members show a single line, identified as \Ha\ by its trace and narrowband excess and tested against \OIII\ and \Hb\ by its partner lines (Appendix~\ref{app:atlas}).
Of the \SuccCandN\ HAE candidates, \SuccCandHa\ are members, a success rate of \SuccCandSuccess\% (68\% interval \SuccCandSuccessLo--\SuccCandSuccessHi\%), and the other \SuccMembersOutN\ members are filler targets.

We name the 11 quasars of the overdensity CH-Q01 to CH-Q11, as CHEF~I does.
We adopt their redshifts from CHEF~I (Appendix~\ref{app:members}).
\Numword{\QsoNCo} have systemic CO(4--3) redshifts from the Atacama Large Millimeter/submillimeter Array (ALMA), to be presented with the ALMA data (program 2025.1.01243.S) in a CHEF paper in preparation.
\Numword{\QsoNMosfire} others have MOSFIRE \Ha\ redshifts, and the remaining \numword{\QsoNSdss} have Dark Energy Spectroscopic Instrument (DESI) Data Release~1 \citep{DESI2025} or SDSS DR16Q \citep{Lyke2020} redshifts ($\pm\QsoSdssErr~\kms$).
The quasars are excluded from the fits.
CHEF~I associates three further quasars with the 11 by friends-of-friends in comoving space, CH-Q12 to CH-Q14, which have DESI redshifts only.
We show them, but the tests use only the 11 that define the overdensity.

\section{Two kinematically cold groups}\label{sec:results}

\subsection{Two velocity groups and a gap}\label{sec:mixture-results}

\begin{figure*}[t]
\centering
\includegraphics[width=\textwidth]{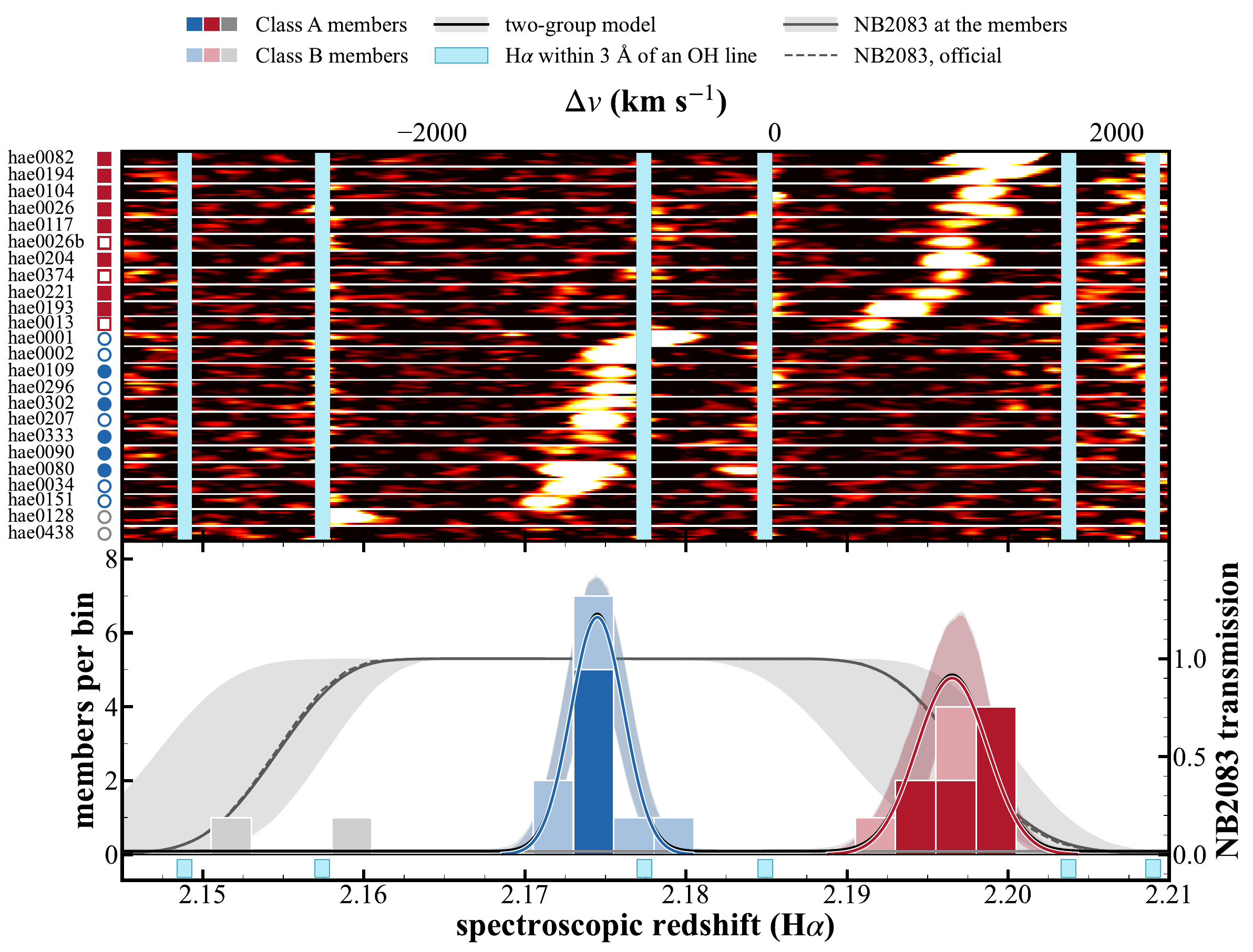}
\caption{Spectra and redshifts of the members.
Top: the central $1\farcs4$ of the two-dimensional spectrum of each member around its trace, minus a running median along wavelength, in units of its robust scatter (color: S/N from 0 to 5).
The strips are sorted by velocity and placed on the \Ha-redshift scale of the bottom axis.
Light blue, also in the boxes under the histogram: within 3~\AA\ of an OH line.
Left: the HAE identifier (Table~\ref{tab:members}), the group (color, gray for the uniform background), and the class (filled: A, open: B).
\NII\ lies at $-670$ and $+945~\kms$ from \Ha, so the \NII~$\lambda6585$ lines of blue members fall in the gap.
Bottom: the members in bins of $\Delta z=0.0025$ (Class A dark, B light) and the two-group model with its three terms and 68\% bootstrap bands.
Right axis: the NB2083 passband at the members (median and range) and the official curve (dashed).
Top axis: $\Delta v$.}
\label{fig:redshifts}
\end{figure*}

The 24 member redshifts form two narrow groups separated by a gap (Figure~\ref{fig:redshifts}).
Thirteen members lie at $\Delta v\le\GapLo~\kms$, two of them isolated at $z=2.152$ and 2.159, and 11 lie at $\Delta v\ge\GapHi~\kms$, so that none lies in the \GapObs~$\kms$ between them.
A mixture of two Gaussian components and a uniform background, a flat term for the members outside both groups (Appendix~\ref{app:members}), places the groups at $z=\BlueRedshift$ and $\RedRedshift$, \PeakSeparation~$\kms$ apart (68\% interval \PeakSeparationLo--\PeakSeparationHi~$\kms$), with equal observed weights.
The gapper estimator of the dispersion, suited to small samples \citep{Beers1990}, gives \RefBlueGapper\ (68\% interval \RefBlueGapperLo--\RefBlueGapperHi) and \RefRedGapper\ (\RefRedGapperLo--\RefRedGapperHi)~$\kms$ for the two groups.
Both groups are kinematically cold, with dispersions below those reported for SSA22, the Spiderweb, and BOSS1244 (350--540~$\kms$) and for the density peaks of Hyperion (320--731~$\kms$), and among the structures of Table~\ref{tab:literature} comparable only to the BOSS1542 filament (Figure~\ref{fig:phase}b).
Groups as cold have been found in a sub-structure of BOSS1244 ($152\pm58~\kms$; \citealt{Shi2021}) and around the radio galaxy MRC~0052$-$241 at $z=2.86$, where two groups of 12 and 19 Ly$\alpha$ emitters have dispersions of 185 and 230~$\kms$ \citep{Venemans2007}.

A Gaussian velocity distribution with the mean and spread of the members (standard deviation \GapSD~$\kms$) would place about \numword{\GapExpectedNone} of them in the gap.
Drawn through the position-dependent narrowband selection (Section~\ref{sec:selection-results}), such a Gaussian leaves so wide a central gap in \numword{\GapNPos} of \GapNSimCases\ samples of 24 (\GapNNone\ without the selection), and no flatter or more peaked distribution does so in more than \GapNShapeMaxPos\ (Appendix~\ref{app:gap}).
The member redshifts are therefore unlikely to come from one single-peaked velocity distribution.
The test does not distinguish two separate structures from one protocluster made of two substructures, which Section~\ref{sec:nature} relates through the mean overdensity of the region around them.

\subsection{The narrowband selection does not create the groups}\label{sec:selection-results}

At the positions of the 11 red members, the red half-maximum edge of the NB2083 passband lies at \Ha\ redshifts of $z=\PosZRedEdgeMin$--$\PosZRedEdgeMax$, which cuts through the red group.
Their \Ha\ lines fall at \PosTRedMin--\PosTRedMax\% of the peak transmission (median \PosTRedMedian\%), whereas the blue members lie on the plateau, the flat top of the passband (Figure~\ref{fig:redshifts}).
The narrowband fluxes of the members show this dependence on position, so we model the selection of each member with the passband at its own position (the per-position passband, Appendix~\ref{app:response}).
Two Gaussian components then describe the members better than one, with a difference of the Bayesian information criterion of $\Delta{\rm BIC}=\PosDeltaBIC$, and of at least \PosDeltaBICMin\ for the other model choices tested (Appendix~\ref{app:response}).
The red group is intrinsically \PosRatio\ times as rich as the blue one (68\% interval \PosRatioLo--\PosRatioHi), and the intrinsic dispersions are \PosDispBlue\ and \PosDispRed~$\kms$.
The selection makes so wide a gap rarer (Section~\ref{sec:mixture-results}), so it does not create the gap.
OH lines do not shape the distribution either, because only \GapOHPercent\% of the gap lies within $\pm3$~\AA\ of an OH line.
The two groups and the empty gap between them therefore belong to the galaxies, not to the way they were selected or observed.

\subsection{Cold but extended groups}\label{sec:extent}

Within the masks the two groups occupy the same area (Figure~\ref{fig:projected}c), and their centroids, weighted by membership probability, lie \ProjectedSeparation~proper Mpc apart (68\% interval \RefProjSepLo--\RefProjSepHi).
Random reassignments of the members give a larger separation with probability \CentroidPValue.
Each group is kinematically cold but extended on the sky.
Its members lie at an rms projected radius of \ExtentRrmsBlue\ (blue, 68\% interval \ExtentRrmsBlueLo--\ExtentRrmsBlueHi) and \ExtentRrmsRed\ (red, \ExtentRrmsRedLo--\ExtentRrmsRedHi) proper Mpc from its centroid, with pairs up to \ExtentPairMaxBlue\ and \ExtentPairMaxRed~proper Mpc apart, so each group stays this cold across the whole section that the masks cover, a lower limit on its extent.
A virialized halo with the dispersion of either group would have a mass of $\RefMassBlueGapper\times10^{12}$ or $\RefMassRedGapper\times10^{12}~\Msun$ for the relation of \citet{Evrard2008} and a radius $r_{200}$, within which the mean density is 200 times the critical density, of only \ExtentRvirBlue\ or \ExtentRvirRed~proper Mpc, smaller than the rms radii by factors of \ExtentRatioBlue\ and \ExtentRatioRed.
The crossing times, the rms radius divided by the dispersion, are \ExtentTcrossBlue\ (\ExtentTcrossBlueLo--\ExtentTcrossBlueHi) and \ExtentTcrossRed\ (\ExtentTcrossRedLo--\ExtentTcrossRedHi)~Gyr, longer than the \ExtentAgeZRef~Gyr age of the Universe at $z_{\rm ref}$, so neither group can be virialized.
With the three-dimensional dispersion $\sqrt3\sigma$ they are still \ExtentTcrossThreeDBlue\ and \ExtentTcrossThreeDRed~Gyr.
A galaxy that moves at the dispersion of its group covers only \ExtentReachBlue\ (blue) and \ExtentReachRed~proper Mpc (red) in the age of the Universe at $z_{\rm ref}$, a small part of the extent of the group (Figure~\ref{fig:phase}).
The dynamical classifications of the cosmic web count an axis as collapsed when its collapse time is shorter than the age of the Universe \citep{ForeroRomero2009}, so neither group has collapsed across the sky, and neither is a knot of the cosmic web.
In pure Hubble flow the dispersions would correspond to rms depths of only \ExtentLosDepthBlue\ and \ExtentLosDepthRed~proper Mpc along the line of sight, and coherent infall would make a structure look thinner still in redshift space \citep{Kaiser1987}.

Simulated protoclusters at $z=2$ of the same extent have dispersions two to three times larger \citep{Chiang2013}, and only the BOSS1542 filament is as cold over a comparable extent (Figure~\ref{fig:phase}b).
The extents of both groups are lower limits, because the masks bound them.
If instead the two groups are parts of one structure that the masks sample only in part, their velocity difference becomes its dispersion, \ExtentSigmaBoth~$\kms$ for the 22 members over an rms radius of at least \ExtentRrmsBoth~proper Mpc, as large as that of the Spiderweb as a whole \citep[$1013\pm87~\kms$;][]{Kuiper2011}.

For each group taken alone, two interpretations fit these numbers.
Each group may be a filament (or sheet) of the cosmic web seen roughly side-on and compressed along the line of sight, as sheets are when they form \citep{Zeldovich1970}, like the BOSS1542 filament at $z=2.24$ with its dispersion of 247~$\kms$ \citep{Shi2021}.
Each group may also be part of an extended protocluster that has not yet collapsed.
In simulations, protoclusters at $z=2$ spread 90\% of their mass over about $35~h^{-1}$~cMpc, and their main halos hold less than 20\% of their galaxies \citep{Muldrew2015}.
The galaxies of the least massive simulated protoclusters have line-of-sight dispersions of $250\pm60~\kms$ at $z\approx3$ and 20--40\% more at $z=2$ \citep{Chiang2013}, somewhat above those of the CH groups but measured over whole protoclusters.
In both interpretations the dispersion measures the depth and internal motions of a structure whose galaxies are not relaxed in its potential \citep{Eisenstein1997,CroneOdekon2024}, and the present data, which end at the edges of the masks, do not distinguish the two.
The two interpretations overlap, because at $z\approx2$ filaments and sheets dominate the cosmic web \citep{Cautun2014} and most galaxies of a protocluster lie outside its main halo.
They differ in whether a main halo dominates the region around each group, and so make different predictions at larger radii.
If each group is a filament (sheet), the narrow velocity band of Figure~\ref{fig:phase} continues beyond the masks, whereas the velocity spread of a protocluster region changes with radius as its galaxies fall toward its main halo.

\subsection{The quasars divide at the same velocities}\label{sec:spatial-results}

\begin{figure*}[t]
\centering
\includegraphics[width=\textwidth]{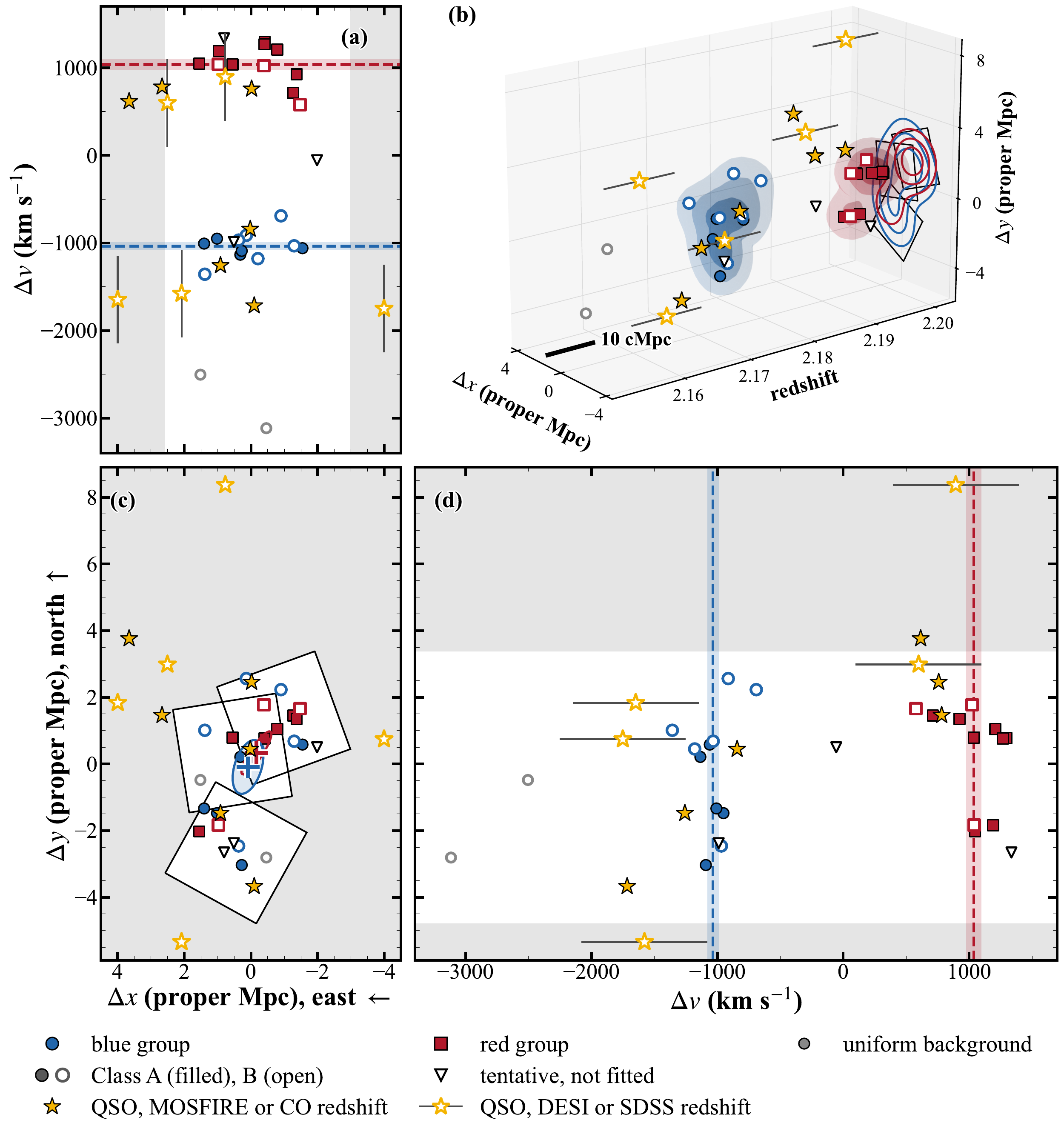}
\caption{The members and the 11 quasars in three orthogonal views and in three dimensions.
Blue circles and red squares: members of the two groups (filled: Class A, open: Class B).
Gray circles: the two isolated members (uniform background).
Triangles: tentative emitters, excluded from the group fits.
Stars: quasars with MOSFIRE or CO (filled) or DESI or SDSS (open, $\pm\QsoSdssErr~\kms$ bars) redshifts.
(a) Velocity against east--west offset from the member centroid, seen from the north of (c).
(b) Redshift space with all three axes on one comoving scale (bar: 10~cMpc), with nested surfaces at 30, 55, and 80\% of the peak kernel density of each group (3.5~cMpc kernel).
The far wall shows the masks and the projected densities at the same fractions.
(c) The sky, east to the left, with the masks and the group centroids (crosses) in their 68\% bootstrap ellipses.
(d) North--south offset against velocity, seen from the west of (c).
Dashed lines and bands: mean velocities with 68\% intervals.
Gray: outside the masks in (c) and offsets covered by no mask in (a) and (d).}
\label{fig:projected}
\end{figure*}

The 11 quasars that define the overdensity divide at the same velocities as the members (Figure~\ref{fig:projected}).
Six lie on the blue side of $z_{\rm ref}$, at a mean velocity of \QsoBlueSideMeanV~$\kms$, and five on the red side, at \QsoRedSideMeanV~$\kms$, with none between \QsoGapLoV\ and \QsoGapHiV~$\kms$.
The red side has the more precise redshifts, because the ALMA setting used here covers the redshift range of the red group and gives CO redshifts for \numword{\QsoNCo} of its quasars.
Their spread of \ExtentQsoCoRange~$\kms$ is therefore limited by that setting and does not measure the coldness of the red structure beyond the members.
On average the red-side quasars lie within \QsoRedSideMinusComponent~$\kms$ of the red group, and the blue-side quasars \QsoBlueSideMinusComponent~$\kms$ blueward of the blue group, where five of the six have DESI or broad-line redshifts with uncertainties of 427--500~$\kms$ (Table~\ref{tab:qsos}).
The \AssocQsoN\ associated quasars of CHEF~I also lie on the red side, at \AssocQsoVMin\ to \AssocQsoVMax~$\kms$ (Figure~\ref{fig:phase}).
Their DESI redshifts come from broad ultraviolet lines, and that of CH-Q09 lies \QsoDesiCoOffsetHi~$\kms$ above its CO redshift, so their offsets of \AssocQsoBeyondRedMin--\AssocQsoBeyondRedMax~$\kms$ from the red group are uncertain.
The red side lies \QsoDecOffset~proper Mpc north of the blue side, and random side labels give a larger centroid separation than the observed \QsoSep~proper Mpc with probability \QsoSepP.
Leaving out any one quasar keeps a gap of at least \SensQsoGapLooMin~$\kms$ (probability $\le\SensQsoGapPLooMax$) and a separation probability of \SensQsoSepPLooMin--\SensQsoSepPLooMax.
With the DESI redshift of the broad absorption-line quasar CH-Q08 instead of its DR16Q redshift, the gap narrows to \SensQsoGapQEightDesi~$\kms$ (probability \SensQsoGapPQEightDesi), and the separation probability stays at \SensQsoSepPQEightDesi.
The quasars reach \QsoRMaxSdss~proper Mpc from the member centroid, and the members about 3~proper Mpc.
The quasars thus carry the division of velocities from the core to the whole overdensity.

\section{The configuration of the two groups}\label{sec:nature}

Two groups \PeakSeparation~$\kms$ apart that overlap on the sky can arise in three configurations.
The members could belong to one virialized halo, or to two halos that fall toward each other within a larger overdense region, as \citet{Kuiper2011} proposed for the Spiderweb, where a larger sample of \Ha\ emitters shows filaments that cross the center instead \citep{Shimakawa2014}.
They could also belong to two structures at different distances along the line of sight, which is how simulated protoclusters matched to SSA22 reproduce its two velocity peaks \citep{Topping2018}.
A spherical model relates the last two through one quantity, the mean matter overdensity of the region that contains both groups (Sections~\ref{sec:boundpair} and \ref{sec:twostructures}).

\subsection{One virialized halo}\label{sec:onehalo}

\begin{figure*}[t]
\centering
\includegraphics[width=\textwidth]{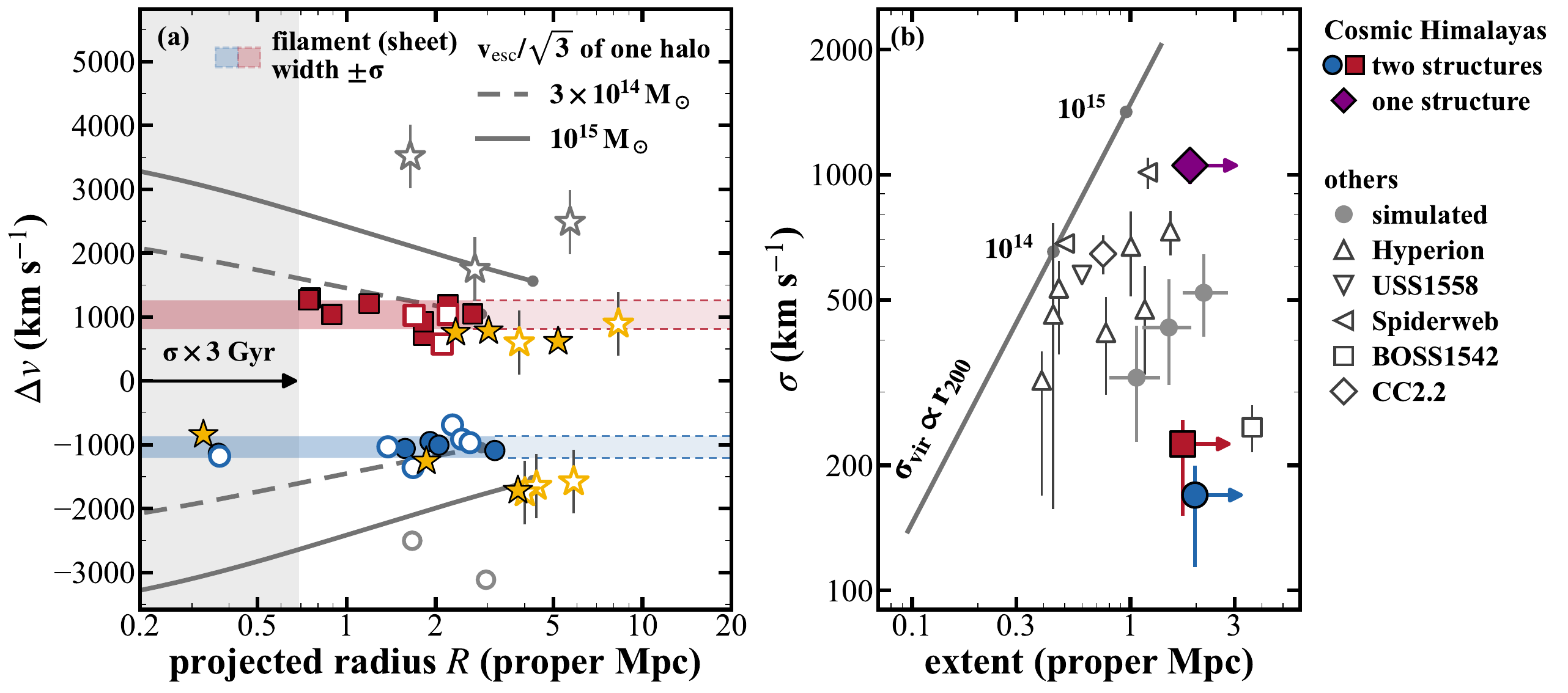}
\caption{(a) Projected phase space of the members and the quasars, relative to $z_{\rm ref}$ and the member centroid, with symbols as in Figure~\ref{fig:projected}.
Open gray stars: the three quasars associated with the 11 in CHEF~I (CH-Q12 to CH-Q14, DESI redshifts, $\pm\QsoSdssErr~\kms$ bars).
Gray curves: the projected escape envelope $v_{\rm esc}/\sqrt3$ of one Navarro--Frenk--White (NFW) halo \citep{Navarro1997} of $M_{200}=3\times10^{14}$ (dashed) or $10^{15}~\Msun$ (solid), with concentration $c=4$, each ending at its turnaround radius (dot).
Colored bands: the mean velocity of each group plus or minus its dispersion $\sigma$, filled over the radii of its members and lighter beyond, where the band of each group would continue as the velocity width of a filament (sheet) seen side-on.
Arrow and gray area: the distance $\sigma\times t_{\rm age}$ that a galaxy moving at the dispersion of the red group covers in the age $t_{\rm age}=\ExtentAgeZRef$~Gyr of the Universe at $z_{\rm ref}$ (\ExtentReachRed~proper Mpc, and \ExtentReachBlue~proper Mpc for the blue group).
(b) Velocity dispersion against extent.
Blue circle and red square: the Cosmic Himalayas as two structures, the gapper dispersion of each group with its 68\% interval.
Purple diamond: the Cosmic Himalayas as one structure, the gapper dispersion of the 22 members of both groups.
The extent of the Cosmic Himalayas is the rms projected radius of the members about their centroid, a lower limit set by the masks (arrows).
Gray line: virialized halos, the dark-matter dispersion of the \citet{Evrard2008} relation against $r_{200}$ at $z_{\rm ref}$, which follows $\sigma\simeq1480~\kms~(r_{200}/{\rm pMpc})^{1.01}$, with dots at $10^{14}$ and $10^{15}~\Msun$.
A structure below the line is larger than $r_{200}$ of the halo that its dispersion implies.
Open symbols: published dispersions and extents of groups and density peaks at $z=2.2$--2.5, one symbol per structure, namely the projected rms radii of the seven density peaks of Hyperion \citep{Cucciati2018}, the projected radius that holds 40\% of the members of CC2.2 \citep{Darvish2020}, half the length of the BOSS1542 filament \citep{Shi2021}, the radii of the richest clump of USS1558$-$003 and of the core group of the Spiderweb \citep{Shimakawa2014}, and the virial radius of the Spiderweb as a whole \citep{Kuiper2011}, so that the Spiderweb appears twice, as its core and as a whole.
Filled gray circles: simulated protoclusters at $z=2$ with Fornax-, Virgo-, and Coma-type descendants (from small to large extent) at their effective radii, with 1.3 times the dispersions at $z\approx3$ (bars: 1.2--1.4 times), as \citet{Chiang2013} find them 20--40\% higher at $z=2$.}
\label{fig:phase}
\end{figure*}

In projected phase space, a virialized halo fills the region inside its escape velocity, concentrated toward small velocities.
The members instead lie at the two velocities of the groups at every projected radius (Figure~\ref{fig:phase}).
Taken as one halo, they would have a virial mass of $\RefMassAllGapper\times10^{14}~\Msun$ (68\% interval \RefMassAllGapperLo--\RefMassAllGapperHi), and \RefNInsideProjectedMid\ of them lie within $1/\sqrt3$ of the escape velocity of a $3\times10^{14}~\Msun$ halo, the range that contains most line-of-sight velocities of bound galaxies.
They do not fill its envelope, however, because the gap between the groups is empty.
\Numword{\QsoNBeyondTurnaroundHigh} of the 11 quasars lie beyond the turnaround radius of even a $10^{15}~\Msun$ halo (\PhaseRtaHigh~proper Mpc), where matter still expands, and lie on the two sides of the same gap.
One virialized halo would therefore need an empty center in its velocity distribution, which single-peaked distributions leave in at most \GapNShapeMaxPos\ of \GapNSimCases\ samples (Section~\ref{sec:mixture-results}), and would have to reach beyond its turnaround radius to hold the quasars.
They could still be the near and far sides of one collapsing region, which coherent infall can split into two velocity peaks \citep{Kaiser1987}, and Section~\ref{sec:twostructures} places this case at one end of the overdensity axis of a spherical model.

\subsection{A spherical model of the pair}\label{sec:boundpair}

\begin{figure}[tb]
\centering
\includegraphics[width=\columnwidth]{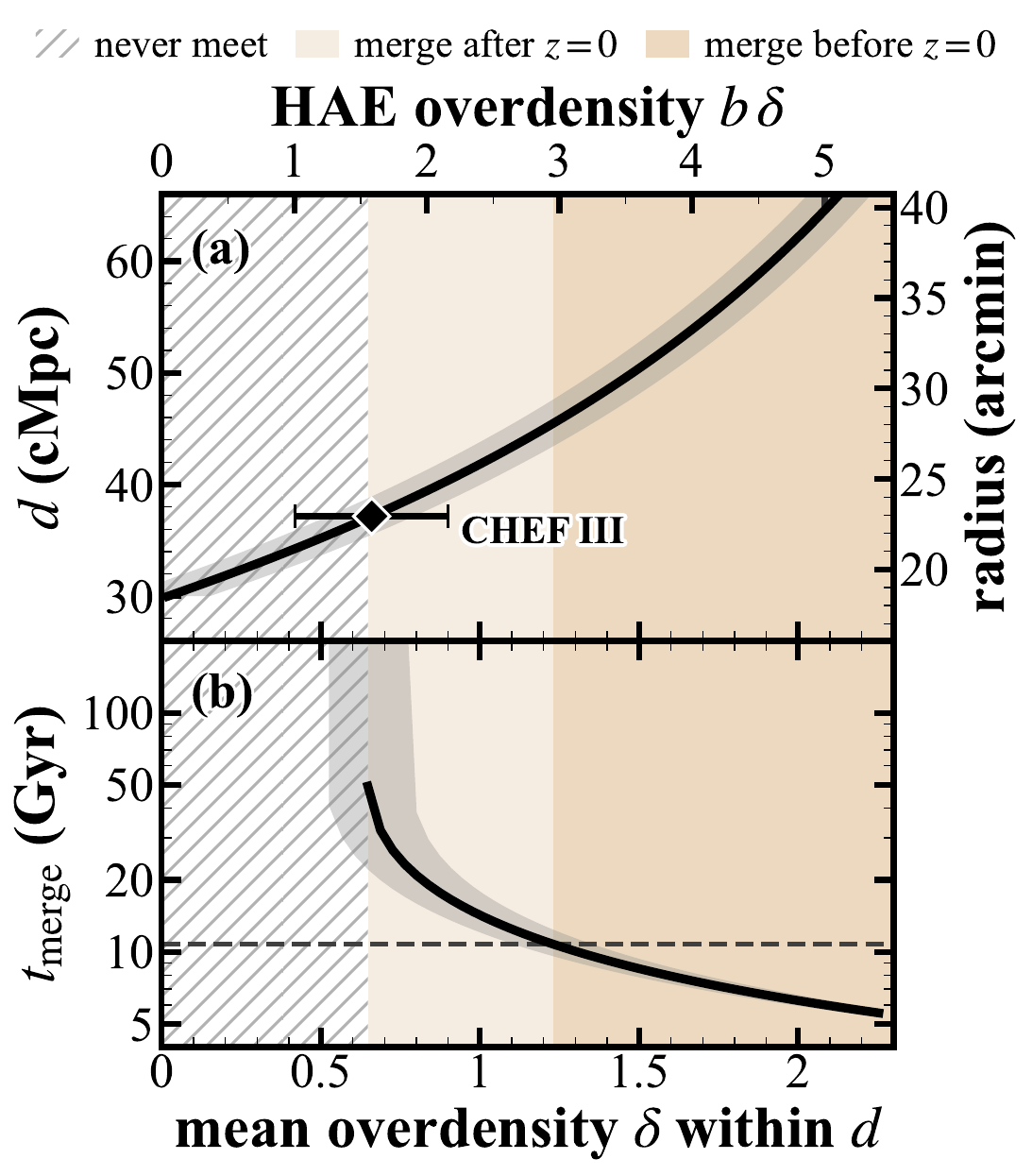}
\caption{The separation and fate of the two structures against the mean overdensity $\delta$ within their separation $d$, in the spherical model (Appendix~\ref{app:spherical}).
Top axis: the HAE overdensity $b\delta$ ($b=2.4$, \citealt{Geach2012,Cochrane2017}).
Gray bands: 68\% ranges over draws of the velocity difference and projected separation.
(a) $d$ for a growing separation, and $d$ as an angle on the sky (right axis).
(b) Time from $z_{\rm ref}$ until $d$ falls to 5\% of its value (band open to the top where more than 16\% of the draws never merge), and the present day (dashed).
Shading: the structures never meet (hatched), merge after $z=0$ (light), or merge before $z=0$ (dark).
Diamond in (a): the range of the preliminary HAE overdensities of CHEF~III over the whole volumes of their maps, from \ChefThreeDmWide\ (NEWFIRM mosaic) to \ChefThreeDmCore\ (MOIRCS core), drawn on the curve at the center of the range.}
\label{fig:fate}
\end{figure}

We describe the physical separation $d$ of the two groups with a spherical model (Appendix~\ref{app:spherical}), which places one group at the center of a sphere and the other on its surface and lets $d$ evolve since the Big Bang under the mass inside the sphere,
\begin{equation}
\ddot d=-\frac{GM(<d)}{d^{2}},\qquad M(<d)=(1+\delta)~\frac{H^{2}d^{3}}{2G},
\label{eq:shell}
\end{equation}
where $\delta$ is the mean matter overdensity within $d$ and $H^2d^3/(2G)$ the mass of the sphere at the mean density.
Because the separation has grown from zero over the age of the Universe, $\delta$ fixes its rate of change $\dot d$ at each $d$, and we observe
\begin{equation}
V_r=\dot d\cos\theta,\qquad \sin\theta=R_p/d,
\label{eq:project}
\end{equation}
along the line of sight, where $R_p$ is the projected separation of the member centroids.
The model omits tangential motion, which would make the pair harder to bind and slower to merge, and matter that streams into the region after $z_{\rm ref}$, which would make it easier to bind and faster to merge.
At the mean density ($\delta=0$) the velocity difference is pure Hubble flow, $\dot d=Hd$, and the groups are \MsepDZero~cMpc apart.
Matter between them slows the recession, to $\dot d\approx Hd~(1-f\delta/3)$ in linear theory with $f=\Omega_m(z)^{0.55}\approx0.96$, so the same velocity difference implies a larger separation, $d\approx V_r/[H(1-f\delta/3)]$ for $\delta\lesssim1$.
Equations~(\ref{eq:shell}) and (\ref{eq:project}) give \MsepDHalf, \MsepDOne, and \MsepDTwo~cMpc for $\delta=0.5$, 1, and 2 (Figure~\ref{fig:fate}a).
These separations exceed the 10--20~$h^{-1}$~cMpc of the simulated pairs that reproduce SSA22 \citep{Topping2018}.

\subsection{Receding, bound, or approaching}\label{sec:twostructures}

The same overdensity decides the fate of the pair (Figure~\ref{fig:fate}b and Appendix~\ref{app:spherical}).
The pair is bound if its separation turns around before the cosmological constant, neglected in Equation~(\ref{eq:shell}) and included in the integration forward in time, dominates the expansion.
For $\delta<\MergeBoundDelta$ ($d<\MergeBoundDCMpc$~cMpc) the separation grows for ever, and the groups are two structures along the line of sight that never meet.
For $\delta>\MergeBoundDelta$ the pair is bound, and the groups are two halos or structures within one overdense region that will fall into each other, after $z=0$ for $\delta<\MergeMergedDelta$ and before $z=0$ above it ($d>\MergeMergedDCMpc$~cMpc).
In all three cases the separation still grows at $z_{\rm ref}$, and in each the quasars divide between the groups, six on the blue side and five on the red side.
Other protocluster fields at $z=2.2$--3.1 have mean matter overdensities of 1.3--1.7 within spheres of 13--29~cMpc, and in simulations half of the regions of $(24~{\rm cMpc})^3$ with $\delta>0.36$ at $z=2$ become clusters \citep{Shi2021,Topping2018,Darvish2020,Chiang2013}.
At the end of the overdensity axis, the separation already shrinks at $z_{\rm ref}$.
It can shrink only if the sphere has turned around, which needs a mean density $9\pi^2/16$ times the cosmic mean ($\delta\geq\MsepDeltaTurn$).
With one group at the center and the other on the surface, approaching at the observed rate needs $\MsepApproachMin\times10^{15}~\Msun$ within $\MsepApproachMinD$~proper Mpc, the density of a virialized halo.
If instead both groups fall toward a halo between them, each at half the separation and half the velocity difference, the same timing needs $\InfallHalfMass\times10^{14}~\Msun$ within \InfallHalfD~proper Mpc of that halo.
A spectroscopic survey of the sphere of radius $d$, \MergeBoundRadiusArcmin--\MergeMergedRadiusArcmin\ arcmin on the sky, can measure $\delta$ and tell these cases apart (Section~\ref{sec:quasars}).

\section{Tests of the configurations}\label{sec:quasars}

The 11 quasars of the densest quasar overdensity of SDSS thus share the velocity split of the two galaxy groups, and a single halo would host them only if it reached beyond its turnaround radius.
On average the red-side quasars lie within \QsoRedSideMinusComponent~$\kms$ of the red group, and the blue-side quasars, five of them with redshifts uncertain by 427--500~$\kms$, lie on the blue side of the gap.
If the groups are two structures, the quasars on each side lie, in both interpretations of Section~\ref{sec:extent}, in a structure that has not collapsed.

A spectroscopic survey of about one square degree can measure the mass, geometry, and fate of the two structures.
A wide-field multi-object spectrograph such as the Subaru Prime Focus Spectrograph (PFS), with a field of view 1.3\degr\ across \citep{Takada2014}, covers the sphere of radius $d$ around the structures in one pointing, and Ly$\alpha$ at $z=2.19$ falls at 3880~\AA, within the range of optical spectrographs.
Ly$\alpha$ redshifts of the Ly$\alpha$ emitters of the field \citep{Liang2025} would map the two structures beyond the masks and test the prediction of Figure~\ref{fig:phase} that the velocity band of a filament (sheet) stays as narrow beyond 3~proper Mpc as within the masks, whereas the velocity spread of a protocluster region changes with radius.
They would also give the mean overdensity within $d$, once the counts are corrected for bias and redshift-space distortions (Appendix~\ref{app:spherical}), which determines whether the two structures will merge (Section~\ref{sec:twostructures}).
The narrowband HAEs of CHEF~III (Y.~Liang et al., in preparation) give a first indication.
Over the whole volumes of their maps, their preliminary matter overdensity for $b=2.4$ is $\delta\approx\ChefThreeDmCore$ (68\% interval \ChefThreeDmCoreLo--\ChefThreeDmCoreHi) for the MOIRCS core and \ChefThreeDmWide\ (\ChefThreeDmWideLo--\ChefThreeDmWideHi) for the NEWFIRM mosaic.
These values would place the two groups about \ChefThreeDCore\ and \ChefThreeDWide~cMpc apart, a bound pair that merges after $z=0$ in the first case and two structures that never meet in the second (Figure~\ref{fig:fate}).
The true volume of the structures is not known, however, because their three-dimensional shape is not.
If they fill only part of the volume of a filter map, the overdensity within $d$ is higher than these values, and the redshifts of a wide-field survey would fix the volume and with it the overdensity.
Narrowband imaging in other emission lines at $z=2.19$ with a wide-field near-infrared imager, such as that of ULTIMATE-Subaru \citep{Minowa2024}, would select different galaxy populations of the two structures.
Background galaxies of the same survey would map the intergalactic H~{\sc i} of the structures by Ly$\alpha$ forest tomography \citep{Lee2016}.
Systemic redshifts of all 11 quasars, with an ALMA setting that also covers the blue group, would show how closely the quasars on each side follow their group.
Such a survey would turn the three configurations of Section~\ref{sec:nature} from possibilities into a measurement.

\section{Conclusion}\label{sec:summary}

MOSFIRE spectroscopy of the core of the densest quasar overdensity of SDSS finds two velocity groups of \Ha\ emitters \PeakSeparation~$\kms$ apart, with no member between them, and the 11 quasars divide at the same velocities.
The groups are kinematically cold, with dispersions of \RefBlueGapper\ and \RefRedGapper~$\kms$, yet extend across the masks, with crossing times longer than the age of the Universe at $z=2.19$, and are either filaments (or sheets) seen side-on or uncollapsed regions of protoclusters.
In a spherical model, the mean overdensity $\delta$ of the region around the groups sets their separation, \MsepDZero--\MsepDOne~cMpc for $\delta=0$--1, and their fate, from two structures that never meet ($\delta<\MergeBoundDelta$) to a bound pair that merges before $z=0$ ($\delta>\MergeMergedDelta$), and the core alone does not decide among these cases.
If the structures are two, the quasars of the densest quasar overdensity of SDSS lie in cold, extended structures that have not collapsed.
A spectroscopic survey of about one square degree with a wide-field instrument such as Subaru PFS can map the structures, distinguish a filament (sheet) from a protocluster region, and measure their mean overdensity, which determines whether they will merge.
The narrowband HAEs of CHEF~III suggest $\delta\approx\ChefThreeDmWide$--\ChefThreeDmCore\ over the whole volumes of their maps, on both sides of the threshold of a bound pair (Figure~\ref{fig:fate}a), and higher values if the structures fill only part of these volumes.
Whether the densest quasar overdensity of SDSS consists of two structures, one structure seen in part, or a pair that will merge is a question that one wide-field spectroscopic survey can answer.

\section*{Data availability}
The member and quasar tables of Appendix~\ref{app:members} are available in machine-readable form, as is a catalog of the 58 targets and the two serendipitous \OIII\ emitters, with positions, masks, spectroscopic classes, redshifts with uncertainties, and the S/N of the primary line.
Two tentative redshifts were determined by visual inspection and have no fitted uncertainty, and the MOSFIRE quasar redshifts are the peaks of the broad \Ha\ lines, with systematic uncertainties given in Appendix~\ref{app:members}.
CHEF~I includes these redshifts in its compilation of the spectroscopic redshifts of the field.
The reduced spectra are available from the corresponding author upon reasonable request, and we plan a public release with the full CHEF data set.

\section*{Acknowledgments}
This work was supported by JSPS KAKENHI Grant Numbers JP24K17084, JP25H00663, JP25H00674, and JP22K21349, and by the JSPS Core-to-Core Program, grant number JPJSCCA20210003.
YL acknowledges the International Leading Research program ``Comprehensive understanding of the formation history of structures in the Universe'' and SUPER-IRNET 1.0 and 2.0.
SC gratefully acknowledges support from the Italian Ministry of University and Research (MUR) under the FARE 2020 program CosmicNodes (grant no. R20S99FS3J, CUP H45E21000490001) and from the European Research Council (ERC) Consolidator Grant 864361 (CosmicWeb).
This work is sponsored by the Chinese Academy of Sciences (CAS), through grant Y329011V01 to the CAS South America Center for Astronomy (CASSACA). WW is supported by the CASSACA Key Research Project E52H540101 / E52H540201 / E52H540301 and the NAOC Research Program E5TG540101.
Some of the data presented herein were obtained at Keck Observatory, which is a private 501(c)3 non-profit organization operated as a scientific partnership among the California Institute of Technology, the University of California, and the National Aeronautics and Space Administration.
The Observatory was made possible by the generous financial support of the W.~M.~Keck Foundation.
The authors wish to recognize and acknowledge the very significant cultural role and reverence that the summit of Maunakea has always had within the Native Hawaiian community.
We are most fortunate to have the opportunity to conduct observations from this mountain.
This paper makes use of the following ALMA data: ADS/JAO.ALMA\#2025.1.01243.S.
ALMA is a partnership of ESO (representing its member states), NSF (USA) and NINS (Japan), together with NRC (Canada), NSTC and ASIAA (Taiwan), and KASI (Republic of Korea), in cooperation with the Republic of Chile.
The Joint ALMA Observatory is operated by ESO, AUI/NRAO and NAOJ.
The National Radio Astronomy Observatory is a facility of the National Science Foundation operated under cooperative agreement by Associated Universities, Inc.
This research is based in part on data collected at the Subaru Telescope, which is operated by the National Astronomical Observatory of Japan (NAOJ; program S24B-135), including data of the Hyper Suprime-Cam (HSC) Subaru Strategic Program \citep[PDR3;][]{Aihara2022} retrieved from the HSC data archive system.
The HSC collaboration includes the astronomical communities of Japan and Taiwan and Princeton University; the HSC instrumentation and software were developed by NAOJ, Kavli IPMU, the University of Tokyo, KEK, ASIAA, and Princeton University, and the HSC pipeline uses software developed for the Vera C. Rubin Observatory.
Funding for the Sloan Digital Sky Survey has been provided by the Alfred P. Sloan Foundation, the U.S. Department of Energy Office of Science, the Heising-Simons Foundation, the U.S. National Science Foundation, and the Participating Institutions.
This research used data obtained with the Dark Energy Spectroscopic Instrument (DESI).
DESI is managed by the Lawrence Berkeley National Laboratory and supported by the U.S. Department of Energy Office of Science (Contract No. DE--AC02--05CH11231), the U.S. National Science Foundation, and the DESI Member Institutions.
The DESI collaboration is honored to be permitted to conduct scientific research on I'oligam Du'ag (Kitt Peak), a mountain with particular significance to the Tohono O'odham Nation.
Data analysis was carried out in part on the Multi-wavelength Data Analysis System and with the cooperation of the Center for Computational Astrophysics, both at NAOJ.
The authors used Claude (Anthropic), ChatGPT (OpenAI), and Gemini (Google) to help write the analysis and figure code and to support proofreading and text checks.
All code, results, and text were checked by the authors, who take full responsibility for them.
Access to ChatGPT was provided through the OpenAI ChatGPT for Academic Researchers program.

\facilities{Keck:I (MOSFIRE), Subaru (MOIRCS, HSC), ALMA, Sloan, Mayall (DESI)}

\software{Astropy \citep{Astropy2013,Astropy2018,Astropy2022}, emcee \citep{ForemanMackey2013}, Matplotlib \citep{Hunter2007}, NumPy \citep{Harris2020}, pandas \citep{pandas2020}, SciPy \citep{Virtanen2020}, SExtractor \citep{Bertin1996}, MOSFIRE data reduction pipeline \citep{Konidaris2019}}

\appendix

\section{Data, classification, and the members}\label{app:data}
\setcounter{equation}{0}

\subsection{Imaging, spectroscopy, and line fits}\label{app:imaging}

We imaged the CH core with Subaru/MOIRCS \citep{Ichikawa2006,Suzuki2008} in NB2083 and $K_s$ on 2024 September 21--23 (UT; Subaru program S24B-135; PI: Y.~Liang), with 8.6--9.0~ks per pointing in NB2083 and 2.4--2.7~ks in $K_s$ in four pointings (Figure~\ref{fig:field}a).
The PSFs have FWHMs of $\ChPsfKsArcsec$ ($K_s$) and $\ChPsfNBArcsec$ (NB2083), and the median $5\sigma$ limits in $2''$ apertures are $K_s=\ChDepthKs$~mag and NB2083 $=\ChDepthNB$~mag.
CHEF~I calibrates the photometry against UKIDSS \citep{Lawrence2007}, VHS \citep{McMahon2013}, and NEWFIRM $K_{Xs}$ imaging to 0.03--0.04~mag and selects the emitters.
The excess significance $\Sigma$ compares the narrowband and continuum fluxes with the noise of empty apertures \citep{Bunker1995,Sobral2013}, and the color cut, analogous to the $BzK$ criterion \citep{Daddi2004}, removes sources with $i-K_s<1.6~(g-i)-0.5$ in the HSC-SSP images \citep{Aihara2018,Aihara2022}.
The band also admits \OIII\ and \Hb\ at $z\approx3.1$--3.3, which the color cut does not remove.
CHEF~III describes this selection with its completeness and contamination.

The masks targeted four CH quasars first and then narrowband-excess sources of the preliminary catalog, in order of their colors, known counterparts, excess, and magnitude, and the members follow the three mask footprints.
The three masks were observed on 2025 December 25--28 (UT; Keck program U239; PI: N.~Reddy) at position angles of 99\degr, 61\degr, and 20\degr, in seeing of $0\farcs62$--$1\farcs30$, for 5.4~hr on source in total, at a resolution of $R\approx2700$ (112~$\kms$).
We reduced the data with the MOSFIRE pipeline \citep{Konidaris2019} and extracted each spectrum with an optimal profile \citep{Horne1986}.
Line fluxes are corrected for slit losses with the spatial profile of each emitter \citep[as in][]{Steidel2014}.
The spectra of CH-Q01 and CH-Q02 are also presented by \citet{Travascio2026}.
The \Ha\ model has a common redshift and width for \Ha, \NII~$\lambda\lambda6550,6585$, and \SII~$\lambda\lambda6718,6733$, is sampled with \texttt{emcee} \citep{ForemanMackey2013}, and excludes pixels within $\pm3$~\AA\ of 61 OH lines \citep{Rousselot2000}.

\subsection{Spectroscopic classes}\label{app:atlas}

\begin{figure*}[t]
\centering
\includegraphics[width=0.62\textwidth]{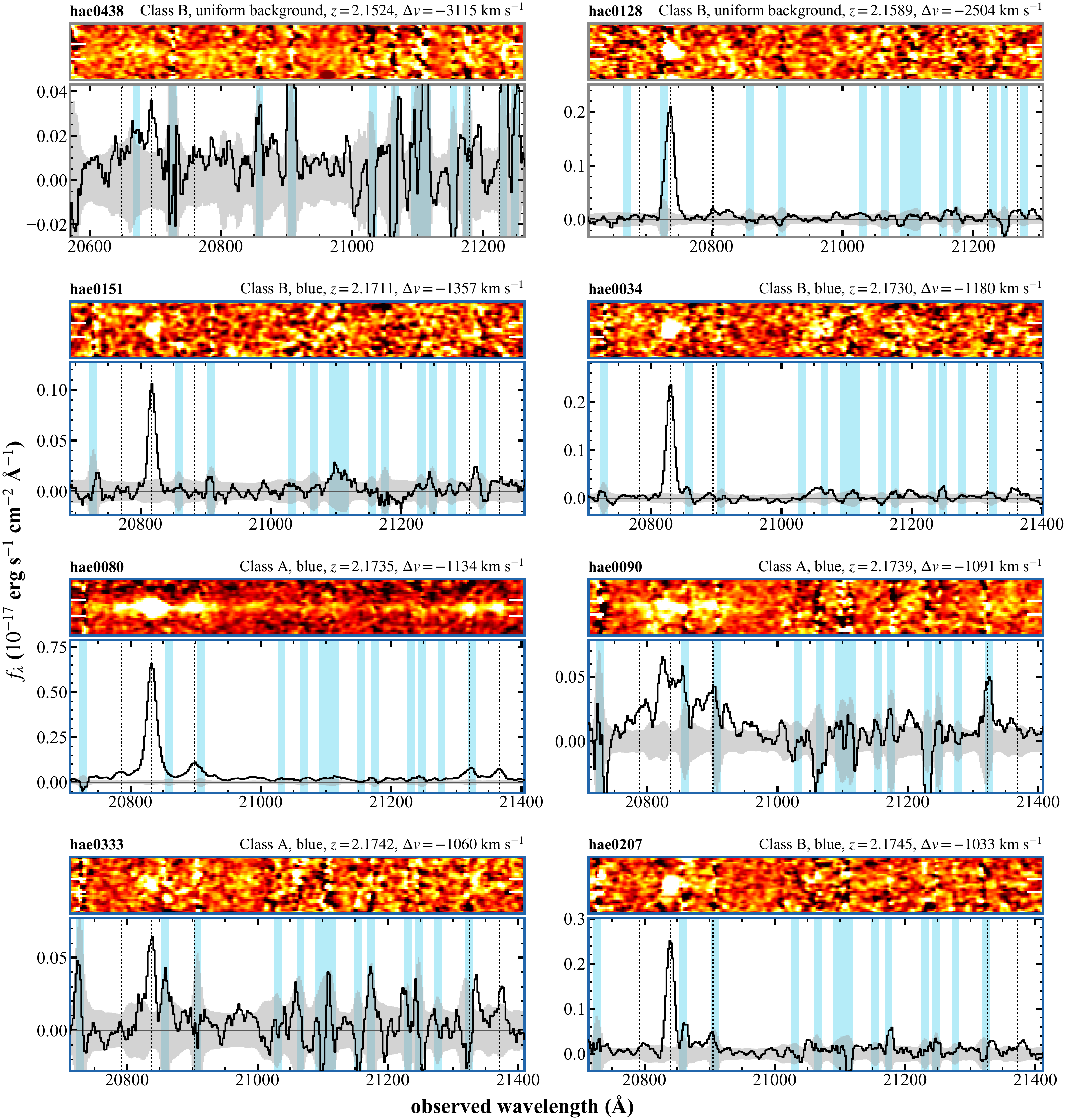}
\caption{Spectral atlas, page 1 of 5, the eight members of lowest velocity. The complete figure set (5 images) is available in the online journal.
Each panel shows the two-dimensional spectrum above the optimal extraction (black, 5-pixel boxcar), with the $\pm1\sigma$ noise as a gray band around zero.
Dotted lines: from left, the fitted positions of \NII~$\lambda6550$, \Ha, \NII~$\lambda6585$, and \SII~$\lambda\lambda6718,6733$.
Light blue: $\pm6$~\AA\ around OH lines.
White ticks: the extraction aperture.
The header gives the class, the group, the redshift, and $\Delta v$.}
\label{fig:atlas}
\end{figure*}

Class A members show \Ha\ with \NII~$\lambda6585$ or \SII\ in the two-dimensional spectrum (primary-line S/N of 8.4--121), and Class B members show a single line, identified as \Ha\ by its spatial trace and narrowband excess (S/N of 3.3--50).
The flux at the position of \OIII~$\lambda5008$ excludes an identification as \OIII~$\lambda4960$ at $3\sigma$ in all 11 Class B members.
The flux at the position of \OIII~$\lambda4960$ excludes \OIII~$\lambda5008$ in seven of them, and in the other four the expected partner, a third of the line flux, would lie below $3\sigma$.
For an identification as \Hb, the $3\sigma$ limit at \OIII~$\lambda5008$ is below 0.6 times the line flux in 10 of the 11 lines, far below the \OIII/\Hb\ ratios of star-forming galaxies at $z\approx2$--3 \citep{Steidel2014}.
Three tentative lines that no fit confirms are not used.
Three targets on mask~2 are \OIII\ or \Hb\ emitters at $z=3.166$, 3.214, and 3.303, and two further \OIII\ emitters, at $z=3.103$ and 3.138, lie on the slits of two targets by chance.
The success rate of the HAE candidates (Section~\ref{sec:data}) combines the purity of the selection with the detectability of the lines and does not measure the completeness, which CHEF~I and CHEF~III measure with injected sources.
No member shows a broad \Ha\ line.
A broad component never both improves the fit by $\Delta{\rm BIC}>10$ and has a FWHM above 1000~$\kms$ \citep{Schneider2010}.
The member hae0090 has $\log(\mbox{\NII}/\mbox{\Ha})=\BroadNinetyNTwo\pm\BroadNinetyNTwoErr$, a ratio indicative of an active nucleus \citep{Strom2017}.
Figure~\ref{fig:atlas} shows the first of the five pages of the spectral atlas, which is available as a Figure Set with the 24 members in order of velocity on three pages, the three tentative lines on the fourth, and the five \OIII\ or \Hb\ emitters on the fifth.

\figsetstart
\figsetnum{6}
\figsettitle{Spectral atlas of the MOSFIRE spectra}

\figsetgrpstart
\figsetgrpnum{6.1}
\figsetgrptitle{Members hae0438 to hae0207}
\figsetplot{fA1_member_atlas_p1.pdf}
\figsetgrpnote{The eight members of lowest velocity. Dotted lines: from left, the fitted positions of \NII~$\lambda6550$, \Ha, \NII~$\lambda6585$, and \SII~$\lambda\lambda6718,6733$. The header gives the class, the group, the redshift, and $\Delta v$. Each panel shows the two-dimensional spectrum above the optimal extraction (black, 5-pixel boxcar), with the $\pm1\sigma$ noise as a gray band around zero. Light blue: $\pm6$~\AA\ around OH lines. White ticks: the extraction aperture.}
\figsetgrpend

\figsetend

\subsection{The members and the two-group model}\label{app:members}

We describe the groups with a mixture of two Gaussian components and a uniform background, fitted by maximum likelihood to the member redshifts $z_i$ and their uncertainties $e_i$ over $2.14<z<2.22$,
\begin{equation}
f(z\mid e_i)=\sum_{k=1}^{2}w_k~\mathcal{N}\!\left(z\mid\mu_k,\sqrt{s_k^2+e_i^2}\right)
+\frac{w_{\rm u}}{\Delta z},
\label{eq:mixture}
\end{equation}
where the weights sum to one, $\Delta z=0.08$, and each width $s_k$ is at least 150~$\kms$.
The uniform background accounts for the two isolated members, and 68\% intervals come from 1000 bootstrap refits.
The 13 Class A members, whose redshifts rest on two or more lines each, show the same two groups, with \numword{\PaperClassABlue} blue and \numword{\PaperClassARed} red members.
The \Ha\ luminosities and \NII/\Ha\ ratios of the members do not differ between the groups (Mann--Whitney $p=\GroupLhaP$ and \GroupNTwoP).
Tables~\ref{tab:members} and~\ref{tab:qsos} list the members, the 11 quasars, and the three associated quasars.
Eleven velocities drawn uniformly over the redshift window of the overdensity leave a gap at least as wide as that of the quasars in a fraction \QsoGapP\ of realizations, and the separation of the two sides of the quasars on the sky persists when each DESI or SDSS redshift is redrawn within its uncertainty (\QsoSepPFracSig\% of the draws give $p<0.05$).

Five of the 11 red members and none of the blue ones lie in close pairs or show two velocity components (Fisher's exact test, $p=\RefPairFisherP$, applied a posteriori).
The pairs are hae0082 and hae0194 ($1\farcs33$ or 11~proper~kpc apart, $\Delta v=29~\kms$) and hae0026 and hae0026b ($1\farcs9$, 150~$\kms$).
The \Ha\ position--velocity diagrams of hae0026 and hae0117 show two components each.
Slits cannot target two galaxies separated across the slit, so these numbers are lower limits.

\begin{deluxetable*}{llllllrrrrrl}
\tabletypesize{\footnotesize}
\tablecaption{The 24 \Ha\ members\label{tab:members}}
\tablehead{
\colhead{ID} & \colhead{R.A.} & \colhead{Decl.} &
\colhead{Class} & \colhead{$z$} & \colhead{$\Delta v$} & \colhead{S/N} &
\colhead{NB2083} & \colhead{$K_s-{\rm NB}$} & \colhead{$p_{\rm blue}$} &
\colhead{$p_{\rm red}$} & \colhead{Notes} \\
& \colhead{(J2000)} & \colhead{(J2000)} & & & \colhead{($\kms$)} & &
\colhead{(AB)} & \colhead{(mag)} & & &
}
\startdata
hae0438 & 02:09:49.69 & +00:44:50.0 & B & 2.15241(23) & $-$3115 & 3.3 & 23.93 & 0.27 & 0.00 & 0.00 & \nodata \\
hae0128 & 02:10:05.24 & +00:49:24.6 & B & 2.15890(5) & $-$2504 & 19.4 & 23.21 & 0.75 & 0.00 & 0.00 & \nodata \\
hae0151 & 02:10:04.20 & +00:52:20.5 & B & 2.17109(5) & $-$1357 & 16.1 & 23.94 & $>$0.89 & 0.89 & 0.00 & \nodata \\
hae0034 & 02:09:51.67 & +00:51:14.4 & B & 2.17298(2) & $-$1180 & 50.4 & 22.78 & 1.40 & 0.98 & 0.00 & \nodata \\
hae0080 & 02:09:55.83 & +00:50:46.2 & A & 2.17346(1) & $-$1134 & 112.3 & 20.77 & 1.34 & 0.98 & 0.00 & \nodata \\
hae0090 & 02:09:55.49 & +00:44:23.5 & A & 2.17392(25) & $-$1091 & 15.8 & 21.67 & 1.01 & 0.99 & 0.00 & \nodata \\
hae0333 & 02:09:41.16 & +00:51:29.9 & A & 2.17425(12) & $-$1060 & 8.4 & 23.12 & 1.05 & 0.99 & 0.00 & \nodata \\
hae0207 & 02:09:43.12 & +00:51:41.8 & B & 2.17453(3) & $-$1033 & 29.9 & 21.56 & 0.67 & 0.99 & 0.00 & \nodata \\
hae0302 & 02:10:04.37 & +00:47:43.8 & A & 2.17481(10) & $-$1007 & 16.7 & 21.55 & 0.98 & 0.99 & 0.00 & \nodata \\
hae0296 & 02:09:56.23 & +00:45:31.1 & B & 2.17522(2) & $-$968 & 30.4 & 22.54 & 1.31 & 0.98 & 0.00 & \nodata \\
hae0109 & 02:10:01.32 & +00:47:27.2 & A & 2.17540(5) & $-$951 & 11.2 & 22.65 & 0.83 & 0.98 & 0.00 & \nodata \\
hae0002 & 02:09:54.41 & +00:55:23.5 & B & 2.17580(4) & $-$913 & 24.6 & 22.06 & 1.62 & 0.98 & 0.00 & \nodata \\
hae0001 & 02:09:46.23 & +00:54:44.3 & B & 2.17817(4) & $-$691 & 39.3 & 20.75 & 1.82 & 0.86 & 0.00 & \nodata \\
hae0013 & 02:09:41.73 & +00:53:37.0 & B & 2.19165(5) & +577 & 17.2 & 22.78 & 1.24 & 0.00 & 0.82 & \nodata \\
hae0193 & 02:09:43.34 & +00:53:13.1 & A & 2.19308(3) & +713 & 44.9 & 20.69 & 0.89 & 0.00 & 0.94 & fg \\
hae0221 & 02:09:42.52 & +00:53:00.8 & A & 2.19533(7) & +924 & 20.1 & 22.37 & 0.86 & 0.00 & 0.98 & \nodata \\
hae0374 & 02:09:50.27 & +00:53:50.6 & B & 2.19639(5) & +1024 & 10.1 & 22.77 & $>$1.27 & 0.00 & 0.98 & \nodata \\
hae0204 & 02:09:57.67 & +00:51:54.2 & A & 2.19653(3) & +1037 & 39.8 & 21.71 & 0.36 & 0.00 & 0.98 & \nodata \\
hae0026b & 02:10:00.98 & +00:46:45.0 & B & 2.19655(4) & +1039 & 24.7 & 22.77 & 1.36 & 0.00 & 0.98 & \nodata \\
hae0117 & 02:10:05.53 & +00:46:22.5 & A & 2.19663(13) & +1047 & 11.4 & 22.68 & 0.40 & 0.00 & 0.98 & kp \\
hae0026 & 02:10:00.86 & +00:46:44.4 & A & 2.19816(7) & +1190 & 28.1 & 22.59 & 1.29 & 0.00 & 0.98 & kp \\
hae0104 & 02:09:47.13 & +00:52:24.7 & A & 2.19835(3) & +1208 & 22.8 & 22.11 & 1.08 & 0.00 & 0.97 & \nodata \\
hae0194 & 02:09:50.11 & +00:51:51.6 & A & 2.19898(7) & +1268 & 36.7 & 20.19 & 1.30 & 0.00 & 0.97 & bl \\
hae0082 & 02:09:50.06 & +00:51:52.7 & A & 2.19929(2) & +1297 & 121.2 & 20.19 & 1.30 & 0.00 & 0.96 & bl \\
\enddata
\tablecomments{$z$: \Ha\ redshift (vacuum, heliocentric) with its uncertainty in units of the last digits.
$\Delta v$: rest-frame velocity relative to $z_{\rm ref}=\PaperZRef$.
S/N: signal-to-noise ratio of the primary line.
NB2083: total magnitude.
$K_s-{\rm NB}$: $1\farcs5$-aperture color after applying the continuum scale, the median NB2083/$K_s$ flux ratio of line-free sources ($>$: lower limit).
$p_{\rm blue}$, $p_{\rm red}$: membership probabilities of the blue and red groups in the mixture fit.
The remainder belongs to the uniform background.
Note kp: two velocity components in the \Ha\ position--velocity diagram (Appendix~\ref{app:members}).
Note bl: blended with another member in the MOIRCS catalog.
Note fg: a galaxy at $z=0.849$ (DESI) lies about $1''$ away and is a separate object in CHEF~I.}
\end{deluxetable*}
\begin{deluxetable*}{llcccclrrrl}
\tabletypesize{\footnotesize}
\tablecolumns{11}
\tablecaption{The quasars of the CH overdensity\label{tab:qsos}}
\tablewidth{0pt}
\tablehead{
\colhead{ID} & \colhead{Designation} & \colhead{$z_{\rm SDSS}$} & \colhead{$z_{\rm DESI}$} &
\colhead{$z_{\rm MOSFIRE}$} & \colhead{$z_{\rm CO}$} & \colhead{Source} & \colhead{$\Delta v$} & \colhead{$\sigma_{\Delta v}$} &
\colhead{$R$} & \colhead{Side} \\
& & & & & & & \colhead{($\kms$)} & \colhead{($\kms$)} & \colhead{(pMpc)} &
}
\startdata
\cutinhead{The 11 quasars of the overdensity}
CH-Q03 & J020921.91+005148.9 & 2.1669 & 2.1669 & \nodata & \nodata & DESI & $-$1750 & 500 & 4.1 & blue \\
CH-Q07 & J020952.54+004308.0 & 2.1715 & 2.1683 & 2.1673 & \nodata & MOSFIRE & $-$1716 & 433 & 3.7 & blue \\
CH-Q08 & J021024.75+005357.7 & 2.1680\tablenotemark{a} & 2.1773 & \nodata & \nodata & DR16Q & $-$1648 & 500 & 4.4 & blue \\
CH-Q04 & J021009.68+003951.5 & 2.1691 & 2.1687 & \nodata & \nodata & DESI & $-$1578 & 500 & 5.7 & blue \\
CH-Q05 & J021000.50+004727.0 & 2.1748 & 2.1729 & 2.1721 & \nodata & MOSFIRE & $-$1258 & 50 & 1.7 & blue \\
CH-Q02 & J020953.48+005113.5 & 2.1907 & 2.1875 & 2.1766 & \nodata & MOSFIRE & $-$843 & 427 & 0.4 & blue \\
CH-Q10 & J021013.02+005613.4 & 2.1997 & 2.1919 & \nodata & \nodata & DESI & +599 & 500 & 3.9 & red \\
CH-Q09 & J021022.06+005745.0 & 2.2000 & 2.2100 & \nodata & 2.1920 & CO & +615 & 52 & 5.2 & red \\
CH-Q01 & J020953.16+005510.9 & 2.1891 & 2.1890 & 2.1931 & 2.1936 & CO & +758 & 52 & 2.5 & red \\
CH-Q11 & J021014.28+005313.5 & 2.1921 & 2.1903 & \nodata & 2.1938 & CO & +782 & 63 & 3.0 & red \\
CH-Q06 & J020959.37+010648.7 & 2.1950\tablenotemark{a} & \nodata & \nodata & \nodata & DR16Q & +893 & 500 & 8.4 & red \\
\cutinhead{Quasars associated with the 11 in CHEF~I}
CH-Q14 & J021014.03+004939.9 & \nodata & 2.2041 & \nodata & \nodata & DESI & +1749 & 500 & 2.7 & red \\
CH-Q12 & J021032.05+005604.2 & \nodata & 2.2119 & \nodata & \nodata & DESI & +2486 & 500 & 5.7 & red \\
CH-Q13 & J021000.59+005312.0 & \nodata & 2.2229 & \nodata & \nodata & DESI & +3514 & 500 & 1.7 & red \\
\enddata
\tablecomments{Ordered by velocity within each part.
ID: name in CHEF~I.
Designation: SDSS for CH-Q01 to CH-Q11, CHEF~I for CH-Q12 to CH-Q14.
$z_{\rm SDSS}$: SDSS DR18 pipeline redshift \citep{Almeida2023}, as used by \citet{Liang2025}.
$z_{\rm DESI}$: DESI DR1 redshift \citep{DESI2025}.
$z_{\rm MOSFIRE}$: peak of the broad \Ha\ line in a decomposition into broad and narrow components.
$z_{\rm CO}$: systemic redshift from the CO(4--3) line observed with ALMA, converted to the heliocentric frame.
Source: the adopted redshift, which is that of CHEF~I, namely $z_{\rm CO}$ where measured, otherwise $z_{\rm MOSFIRE}$, otherwise $z_{\rm DESI}$, otherwise the SDSS DR16Q redshift (DR16Q), except for CH-Q08 (note a).
The DESI redshifts of the three CO quasars differ from the CO redshifts by \QsoDesiCoOffsetLo\ to \QsoDesiCoOffsetHi~$\kms$.
The SDSS redshifts of three of the four quasars observed with MOSFIRE differ from the \Ha\ redshifts by \QsoSdssOffsetLo\ to \QsoSdssOffsetHi~$\kms$, and that of the broad absorption-line quasar CH-Q02 by \QsoSdssOffsetQTwo~$\kms$.
The SDSS redshifts of the three CO quasars differ from the CO redshifts by \QsoCoSdssOffsetLo\ to \QsoCoSdssOffsetHi~$\kms$, and the \Ha\ redshift of CH-Q01 by \QsoMosfireMinusCoQOne~$\kms$.
$\Delta v$: rest-frame velocity relative to $z_{\rm ref}$.
$\sigma_{\Delta v}$: \QsoSdssErr~$\kms$ for SDSS and DESI redshifts.
For \Ha\ redshifts, $\sigma_{\Delta v}$ is the difference between the peak of the broad line and the center of a single-Gaussian fit (up to \QsoBroadPeakMax~$\kms$), but at least 50~$\kms$.
For CO redshifts, it is the uncertainty of the line fit combined with 50~$\kms$.
$R$: projected distance from the member centroid.
Side: side of $z_{\rm ref}$ on which the quasar lies.
CH-Q12 to CH-Q14 join CH-Q01 to CH-Q11 by friends-of-friends in comoving space at a linking length of \AssocQsoLinkLength~cMpc (CHEF~I) and enter none of the tests.}
\tablenotetext{a}{SDSS DR16Q redshift \citep{Lyke2020}, adopted by CHEF~I; the DR18 redshifts are 2.1962 (CH-Q06) and 2.1693 (CH-Q08).
For the broad absorption-line quasar CH-Q08, CHEF~I does not adopt the DESI redshift, whose template fit misplaces the broad lines.}
\end{deluxetable*}

\section{The NB2083 selection and the gap}\label{app:response}
\setcounter{equation}{0}

For member $i$, the per-position passband $T_i$ is the official curve shifted to the angle of incidence of each NB2083 exposure that covers the member and averaged over these exposures, with an effective refractive index of 1.85 (CHEF~I).
For line fluxes with a cumulative distribution $N(>F)\propto F^{-\beta}$, the selected fraction at redshift $z$ scales as $S_i(z)=T_i(z)^{\beta}$, with $T_i$ at the wavelength of \Ha\ relative to its peak.
We adopt $\beta=1$ and test 0.6 and 1.5, the slopes of the \Ha\ luminosity function at $z\approx2.2$ below and above its knee \citep{Sobral2013}.
Member $i$ is then observed at $z_i$ with probability
\begin{equation}
p(z_i)=\frac{S_i(z_i)~f(z_i\mid e_i)}{\int S_i(z)~f(z\mid e_i)~dz},
\label{eq:selmixture}
\end{equation}
so that the weights $w_k$ of Equation~(\ref{eq:mixture}) describe the groups before selection.
Two Gaussian components describe the members better than one by $\Delta{\rm BIC}\geq\PosDeltaBICMin$ for $\beta=0.6$, 1, and 1.5 and for effective indices of 1.5, 1.85, and 2.0.
A forward model that draws galaxies from a Gaussian at the positions of the members of a group and keeps each with probability $S_i$ gives the intrinsic dispersions of Section~\ref{sec:selection-results}.

The ratio of the narrowband line flux of each detection to its slit-corrected spectroscopic flux traces the passband.
Red members near the pointing centers show the narrowband flux expected on the plateau, whereas the two at the largest angles of incidence, hae0204 and hae0117, show less than half of it.
A single passband fitted to all \PaperRespN\ detections (the empirical response) disfavors the official curve by $\Delta\chi^2=\PaperRespDchi$.
The empirical response (FWHM \PosEmpFWHM~\AA) is wider than the passband at any one position and than the average of the per-position passbands over the same detections (FWHM \PosAvgFWHM~\AA).
Position-independent curves bracket the intrinsic ratio of the red to the blue group, \PosRatio\ with the per-position passbands, at \RefRatioEmpirical\ with the empirical response and \RefRatioOfficial\ with the official curve.

\subsection{The gap between the groups}\label{app:gap}

We draw samples of 24 velocities from single-peaked distributions with the mean and standard deviation of the members (\GapSD~$\kms$): a Gaussian and generalized normal distributions with a density proportional to $\exp[-(|v-\bar v|/a)^p]$ for $p=1$, 1.5, 3, 4, and 8 (nearly flat-topped).
Through the per-position passbands, each drawn galaxy is placed at the position of one of the 54 targets other than the quasars and kept with a probability equal to the transmission of the passband there.
In each sample we find the widest gap among members 7 to 18 of the sorted velocities, where the observed gap lies, because the widest gaps of any sample lie in its sparse tails.
For the Gaussian, a gap wider than the observed \GapObs~$\kms$ occurs in \GapNNone\ of \GapNSim\ samples without selection, in \GapNEmp\ with the empirical response, in \numword{\GapNOff} with the official curve, and in \numword{\GapNPos} of \GapNSimCases\ through the per-position passbands.
Every shape leaves so wide a central gap in at most \GapNShapeMaxPos\ of \GapNSimCases\ samples through the per-position passbands.

\section{The spherical model of the separation}\label{app:spherical}
\setcounter{equation}{0}

We treat the separation $d$ of the two structures as the radius of a spherical shell centered on one of them, which evolves under the mass $M(<d)=(1+\delta)\bar M(d)$ that it encloses, where $\bar M=H^2d^3/(2G)$ is the mass within $d$ at the mean density.
With the time since the Big Bang set to $t=2/(3H(z))$ and the cosmological constant (6\% of $H^2$ at $z=2.2$) neglected, a bound shell follows the cycloid $d=A(1-\cos\eta)$, $t=B(\eta-\sin\eta)$, with $A^3=GMB^2$, and $\delta$ fixes $\dot d$ at a given $d$.
We observe the velocity difference $V_r=\dot d\cos\theta$, where $\sin\theta=R_p/d$ for the projected separation $R_p$ of the member centroids.
For each $d>R_p$ we solve for the $\delta$ that gives $V_r$ on the growing branch ($\eta<\pi$, or unbound) and on the shrinking branch ($\pi<\eta<2\pi$), over 400 joint draws of $V_r$ and $R_p$.
The smallest mass on the shrinking branch is the $\MsepApproachMin\times10^{15}~\Msun$ of Section~\ref{sec:twostructures}, for one group at the center and the other on the surface, whereas the point-mass criterion of \citet{BeersGellerHuchra1982}, which ignores the expansion and the time available, requires $M\geq\RefMinBindMass\times10^{14}~\Msun$.

To follow a growing separation into the future, we integrate $\ddot d=-GM(<d)/d^{2}+\Omega_\Lambda H_0^{2}~d$ from $z_{\rm ref}$ with $M(<d)$ fixed.
The pair is bound when it turns around before the cosmological constant dominates, and it has merged when $d$ falls to 5\% of its value at $z_{\rm ref}$.
For merger criteria of \SensFractionMin--\SensFractionMax\% of the separation at $z_{\rm ref}$ the merged threshold stays at the same $\delta$, and with the velocity difference corrected for the selection (\SensDvSel~$\kms$) the thresholds are \SensBoundDeltaSel\ and \SensMergedDeltaSel.
In a configuration that is not spherical, $\delta$ describes the mean overdensity of the matter that slows the growth of the separation.
At the bound threshold the mass within $d$ is $\MergeMassAtBound\times10^{16}~\Msun$, and the rms fluctuations of the linear density field on the two threshold scales are \MergeBoundSigmaLin\ and \MergeMergedSigmaLin.
In linear theory \citep[spherical collapse;][]{Bernardeau1994} the thresholds are then \MergeBoundNSigma$\sigma$ and \MergeMergedNSigma$\sigma$ fluctuations, which a random region of these sizes exceeds with probabilities of \MergeBoundPRandom\ and \MergeMergedPRandom.
CH was selected as the densest concentration of quasars in the SDSS footprint.
For HAEs with a bias of $b=2.4$ \citep{Geach2012,Cochrane2017}, the thresholds correspond to galaxy overdensities of about \MergeBoundDeltaGal\ and \MergeMergedDeltaGal\ in real space, within a sphere that extends over $\pm\MergeBoundDzHalf$--\MergeMergedDzHalf\ in redshift.
A constrained reconstruction of the density and velocity fields \citep{Horowitz2019,Ata2022} would combine the redshifts of a spectroscopic survey with the quasar overdensity and the H~{\sc i} tomography of the field \citep{Liang2025} and would model the redshift-space distortions of coherent infall.

\section{Comparison with other protoclusters}\label{app:comparison}
\setcounter{equation}{0}

Table~\ref{tab:literature} lists protoclusters at $z>2$ with several velocity groups or density peaks.
In simulations matched to SSA22, its double peak arises when a $\sim10^{15}~h^{-1}\Msun$ protocluster lies 10--20~$h^{-1}$~cMpc from a less massive neighbor along the line of sight, and the two remain distinct structures at $z=0$ \citep{Topping2018}.
The two components of the Spiderweb come from a double-Gaussian fit to the velocities of 38 galaxies \citep{Kuiper2011}, and a larger \Ha\ sample shows filaments that cross its center instead \citep{Shimakawa2014}.
MQN01 at $z\approx3.25$ has a core and a candidate infalling group within 1000~$\kms$ of it \citep{Pensabene2024}, inside an elongated overdensity of Lyman-break galaxies \citep{Galbiati2025}.
Its members observed with JWST/NIRSpec are star-forming, except a quiescent galaxy of stellar mass $\sim10^{11}~\Msun$ that lies 48~kpc from the host of an X-ray-detected active galactic nucleus \citep{Wang2026}.
The Cosmic Vine at $z=3.44$ consists of a core and three further subgroups, and a second structure lies about 87~cMpc in front of it along the line of sight \citep{Jin2024,Sillassen2026}, farther than the two CH groups would be apart as separate structures (\MsepDZero--\MsepDOne~cMpc for $\delta=0$--1, Section~\ref{sec:boundpair}).
Of its 11 quiescent galaxies, two form a merging pair in the core, three lie in two dense subgroups, and six lie outside the densest regions \citep{Sillassen2026}.
Different tracers and sample sizes limit the comparison of dispersions, and each CH dispersion rests on 11 members.

\begin{deluxetable*}{lcccccl}
\tablecaption{Protoclusters at $z>2$ with several velocity groups or density peaks\label{tab:literature}}
\tablehead{\colhead{Structure} & \colhead{$z$} & \colhead{$\Delta v$} &
\colhead{$N$} & \colhead{$\sigma_v$} & \colhead{Interpretation} &
\colhead{Reference} \\
& & \colhead{($\kms$)} & & \colhead{($\kms$)} & &}
\startdata
\begin{tabular}[t]{@{}l@{}}Cosmic\\Himalayas\end{tabular} & 2.175, 2.197 & \PeakSeparation & 11, 11 & \RefBlueGapper, \RefRedGapper & three configurations open (Section~\ref{sec:nature}) & this work \\
SSA22 & 3.069, 3.095 & $\approx$1900 & 43, 103 & $350\pm53$, $540\pm40$ & distinct structures to $z=0$ & \citet{Topping2016,Topping2018} \\
Spiderweb & 2.16 & 1600 & 38 in total & 492, 417 & merger of two halos & \citet{Kuiper2011} \\
 & & & & & or filaments crossing the center & \citet{Shimakawa2014} \\
BOSS1244 & 2.230, 2.246 & $\approx$1500 & 46 in field & $405\pm202$, $377\pm99$ & two groups, 21.6~cMpc apart on the sky & \citet{Shi2021} \\
BOSS1542 & 2.241 & \nodata & 36 & $247\pm32$ & filament & \citet{Shi2021} \\
CC2.2 & 2.232 & \nodata & 47 & $645\pm69$ & one structure, not yet virialized & \citet{Darvish2020} \\
Hyperion & 2.42--2.51 & \nodata & 3--29 per peak & 320--731 & seven peaks, proto-supercluster & \citet{Cucciati2018} \\
\enddata
\tablecomments{$z$: redshifts of the groups.
For the Spiderweb, the double-Gaussian fit that gives $\Delta v$ and $\sigma_v$ does not give group redshifts, and $z$ is the mean redshift.
$\Delta v$: their velocity separation.
$N$: spectroscopic members of each group where given.
$\sigma_v$: velocity dispersion of each group as published, from the gapper estimator (Cosmic Himalayas), the standard deviation (SSA22), a double-Gaussian fit (Spiderweb), a Gaussian fit (BOSS1244 and BOSS1542), the clipped standard deviation (CC2.2), and the biweight or gapper estimator per peak (Hyperion).}
\end{deluxetable*}

\bibliography{references}
\bibliographystyle{aasjournal}

\end{document}